\documentclass[10pt,letterpaper]{article}
\usepackage[top=0.85in,left=2.75in,footskip=0.75in]{geometry}

\usepackage{amsmath,amssymb}

\usepackage{changepage}

\usepackage[utf8x]{inputenc}

\usepackage{textcomp,marvosym}

\usepackage{cite}

\usepackage{nameref}

\usepackage[right]{lineno}

\usepackage{microtype}
\DisableLigatures[f]{encoding = *, family = * }

\usepackage[table]{xcolor}

\usepackage{array}

\usepackage{todonotes,bigints,caption,subcaption}
\reversemarginpar

\newcolumntype{+}{!{\vrule width 2pt}}

\newlength\savedwidth

\newcommand\thickhline{\noalign{\global\savedwidth\arrayrulewidth\global\arrayrulewidth 2pt}%
\hline
\noalign{\global\arrayrulewidth\savedwidth}}

\raggedright
\usepackage[aboveskip=1pt,labelfont=bf,labelsep=period,justification=raggedright,singlelinecheck=off]{caption}

\makeatletter
\renewcommand{\@biblabel}[1]{\quad#1.}
\makeatother

\date{}

\usepackage{lastpage,fancyhdr,graphicx}
\usepackage{epstopdf}
\fancyheadoffset[L]{2.25in}
\fancyfootoffset[L]{2.25in}

\newcommand{\citet}[1]{\cite{#1}}
\newcommand{\citep}[1]{\cite{#1}}
\newcommand{\Laurie}{Laurie \textit{et al.}~\cite{Laurie2015}}
\newcommand{\methods}{Materials and Methods}

\begin{document}
\vspace*{0.2in}

\begin{flushleft}
{\Large
\textbf\newline{Sequential infection experiments for quantifying innate and adaptive immunity during influenza infection} 
}
\newline
\\
Ada W. C. Yan\textsuperscript{1,2*},
Sophie G. Zaloumis\textsuperscript{3},
Julie A. Simpson\textsuperscript{3},
James M. McCaw\textsuperscript{1,3,4}
\\
\bigskip
\textbf{1} School of Mathematics and Statistics, The University of Melbourne, Parkville, Victoria, Australia
\\
\textbf{2} MRC Centre for Global Infectious Disease Analysis, Department of Infectious Disease Epidemiology, School of Public Health, Imperial College London, London,
 United Kingdom
\\
\textbf{3} Centre for Epidemiology and Biostatistics, Melbourne School of Population and Global Health, The University of Melbourne, Parkville, Victoria, Australia
\\
\textbf{4} Modelling and Simulation, Infection and Immunity Theme, Murdoch Childrens Research Institute, The Royal Children’s Hospital, Parkville, Victoria, Australia
\\
\bigskip

* a.yan@imperial.ac.uk

\end{flushleft}
\section*{Abstract}

Laboratory models are often used to understand the interaction of related pathogens via host immunity.
For example, recent experiments where ferrets were exposed to two influenza strains within a short period of time have shown how the effects of cross-immunity vary with the time between exposures and the specific strains used.
On the other hand, studies of the workings of different arms of the immune response, and their relative importance, typically use experiments involving a single infection.
However, inferring the relative importance of different immune components from this type of data is challenging.
Using simulations and mathematical modelling, here we investigate whether the sequential infection experiment design can be used not only to determine immune components contributing to cross-protection, but also to gain insight into the immune response during a single infection.

We show that virological data from sequential infection experiments can be used to accurately extract the timing and extent of cross-protection.
Moreover, the broad immune components responsible for such cross-protection can be determined.
Such data can also be used to infer the timing and strength of some immune components in controlling a primary infection, even in the absence of serological data.
By contrast, single infection data cannot be used to reliably recover this information.
Hence, sequential infection data enhances our understanding of the mechanisms underlying the control and resolution of infection, and generates new insight into how previous exposure influences the time course of a subsequent infection.

\section*{Author summary}

The resolution of an influenza infection requires different components of the immune response to work together.
Despite recent advances in mathematical modelling, we do not well understand how much each broad immune component contributes to this process at a given time.
In this study, we show that laboratory data on the amount of virus over the course of a single infection is insufficient for inferring the contribution of each broad immune component.
However, if the animals are exposed to two different virus strains with only days separating exposures, then the timing and strength of protection provided by the first infection against the second provides crucial additional information. 
We show how mathematical models can be used to recover the timing and strength of each immune component, thus enhancing our understanding of how an infection is controlled, and how a previous exposure changes the time course of a subsequent infection.


\section*{Introduction}

The influenza virus infects epithelial cells in the respiratory tract, causing respiratory symptoms such as coughing and sneezing, and systemic symptoms such as fever.
Three main components of the immune response  --- innate, humoral adaptive and cellular adaptive immunity --- work together to control an infection.
Experiments have revealed the contribution of each major immune component to resolution of an infection, by suppressing each immune component in turn~\citep{Dobrovolny2013,Seo2002,Iwasaki1977,Yap1978,Kris1988}. 
However, current mathematical models do not agree on how each major immune component contributes quantitatively.

A study by Dobrovolny \textit{et al.}~\citet{Dobrovolny2013} highlights these discrepancies.
The study showed that eight existing viral dynamics models~\citep{Baccam2006,Handel2010,Pawelek2012,Saenz2010,Hancioglu2007,Bocharov1994,Miao2010,Lee2009a} made different qualitative predictions when different components of the immune response were removed.
Each model failed to reproduce the effect of removing at least one of the three components discussed above.
The discrepancies arose because many models were only fitted to viral load data from a single infection.

It has been shown that many models can fit the viral load for a single infection well, including models without a time-dependent immune response which are thought to be less biologically realistic~\citep{Baccam2006}; however, if data for multiple initial conditions are available, the viral load may have more features to distinguish between competing models~\citep{Li2014,Ahmed2017}.
One way of altering the initial conditions is through a previous or ongoing infection.
We previously conducted a series of experiments where ferrets were sequentially infected with two influenza strains~\citep{Laurie2015,Laurie2018}.
When a short time interval (1--14 days) separated exposures, a primary infection protected against a subsequent infection.
This protection likely arose through cross-immunity, whereby the immune response stimulated by one strain also protects against infection with another.

While immune markers indicated the approximate timing of each arm of the immune response~\cite{Carolan2016}, the strength of cross-protection due to each component was difficult to measure experimentally.
We hypothesised that mathematical models can be used to gain further insight from these types of experiments.
Few existing models include interactions between influenza strains on short timescales; hence, we constructed viral dynamics models to reproduce the qualitative observations of these experiments~\citep{Cao2015,Yan2016}.
The models also reproduce observations from a range of experiments where immune components were suppressed~\citep{Cao2016}.

Here, we use simulations to show that these mathematical models allow us to extract the timing and strength of cross-protection from sequential infection data.
By attributing cross-protection to specific immune components, the models lead to new insight into how previous exposure influences the time course of a subsequent infection.
Moreover, we find that compared to single infection experiments, sequential infection experiments provide richer information on host immunity, and thus are potentially a powerful tool to study immune-mediated control of a primary infection.

\section*{Results}

\subsection*{Synthetic data}

As a first step to compare the information made available by sequential infection versus single infection experiments, we generated synthetic datasets for each scenario.
Mimicking the experimental procedure of \Laurie{}, we generated a sequential infection dataset where ferrets were exposed to two influenza strains, with intervals of 1, 3, 5, 7, 10 and 14 days between exposures; and a single infection dataset where ferrets were exposed once only.
Details are given in the \methods{} section.
Using synthetic data means that we know the `true' contribution of each immune component in resolving a single infection, and the `true' extent of cross-protection between infections.

\begin{figure}[!h]
\begin{adjustwidth}{-2.25in}{0in}
\centering
\begin{subfigure}[t]{.32\textwidth}
\captionsetup{justification=centering}
\includegraphics[width =\textwidth]{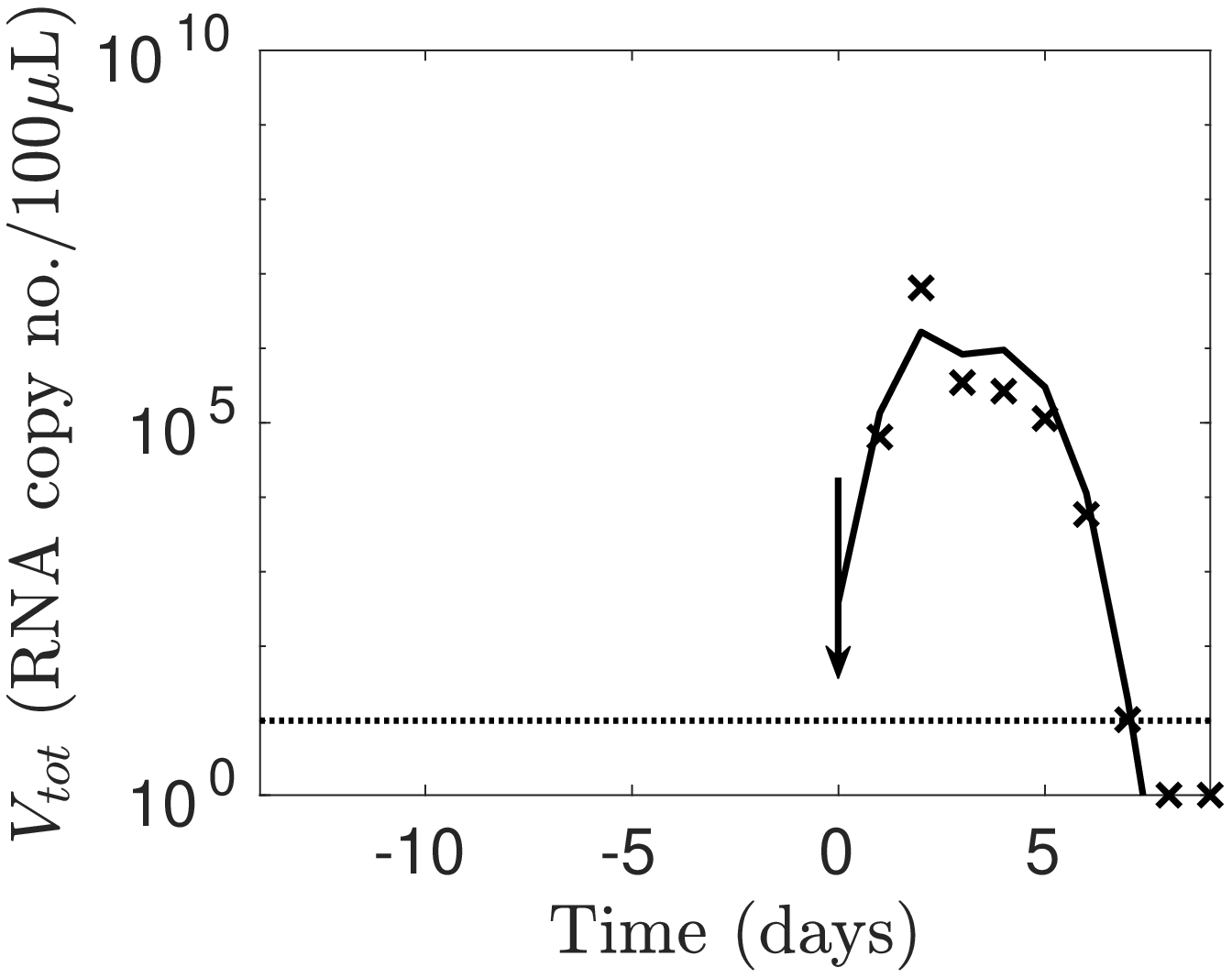}
\caption{single infection}
\end{subfigure}
\begin{subfigure}[t]{.32\textwidth}
\captionsetup{justification=centering}
\includegraphics[width =\textwidth]{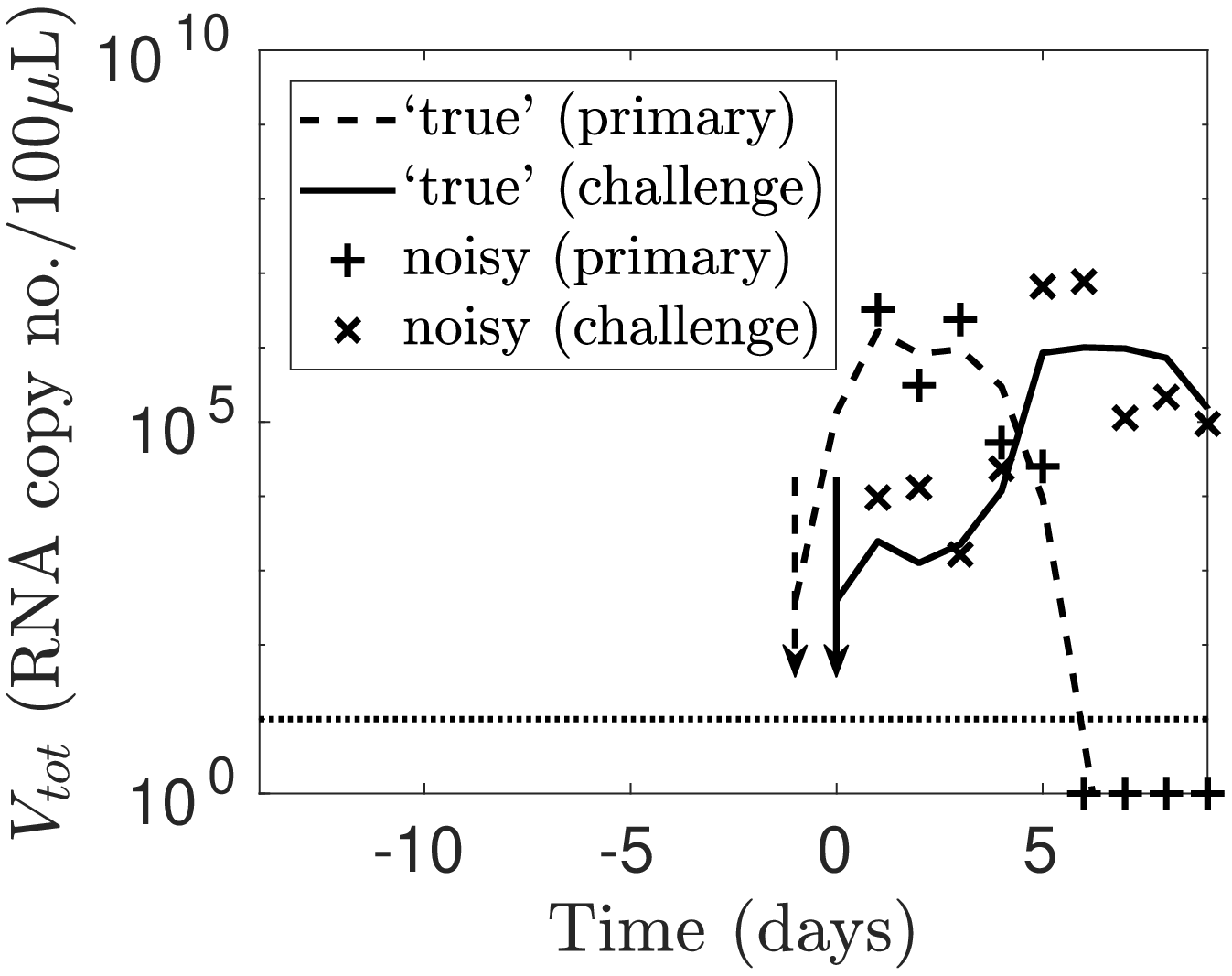}
\caption{1-day interval}
\end{subfigure}
\begin{subfigure}[t]{.32\textwidth}
\captionsetup{justification=centering}
\includegraphics[width =\textwidth]{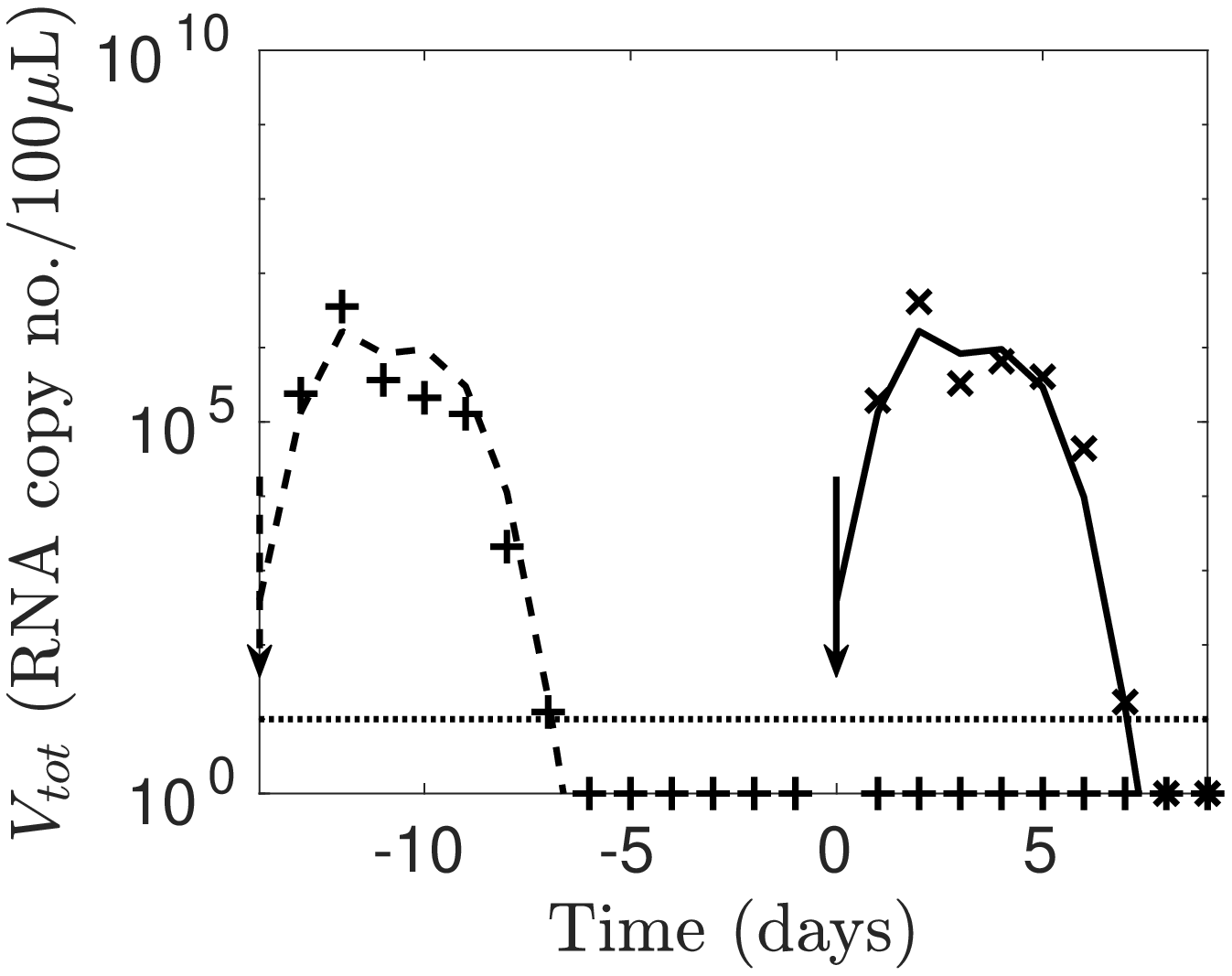}
\caption{14-day interval}
\end{subfigure}
\caption{{\bf A subset of the synthetic data.}
(a) The line shows the simulated `true' viral load for a single infection, with the arrow showing the time of exposure.
The simulated observed viral load is shown as crosses.
The horizontal line indicates the observation threshold (10 RNA copy no./100$\mu$L); observations below this threshold are plotted below this line.
Values below the observation threshold were treated as censored.
(b--c) For sequential infections with the labelled inter-exposure interval, the dashed and dotted lines show the simulated `true' viral load for a primary and challenge infection respectively; the arrows show the times of the primary and challenge exposures.
The simulated observed viral load is shown as crosses.}
\label{fig:data}
\end{adjustwidth}
\end{figure}

\begin{figure}[!h]
\centering
\begin{subfigure}[t]{.32\textwidth}
\captionsetup{justification=centering}
\includegraphics[width =\textwidth]{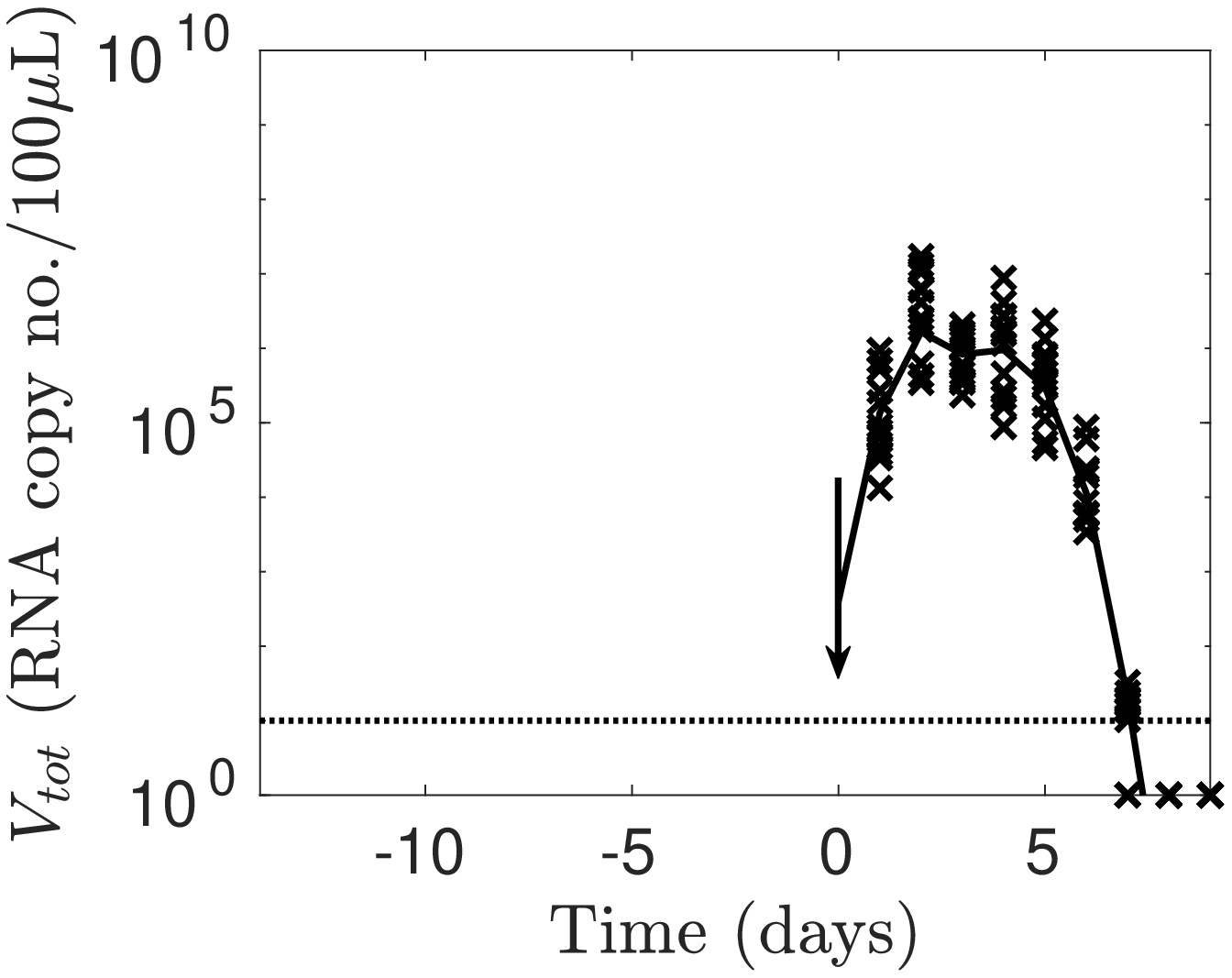}
\caption{single infection}
\end{subfigure}
\begin{subfigure}[t]{.32\textwidth}
\captionsetup{justification=centering}
\includegraphics[width =\textwidth]{figs_main/data_1.eps}
\caption{1-day interval}
\end{subfigure}
\begin{subfigure}[t]{.32\textwidth}
\captionsetup{justification=centering}
\includegraphics[width =\textwidth]{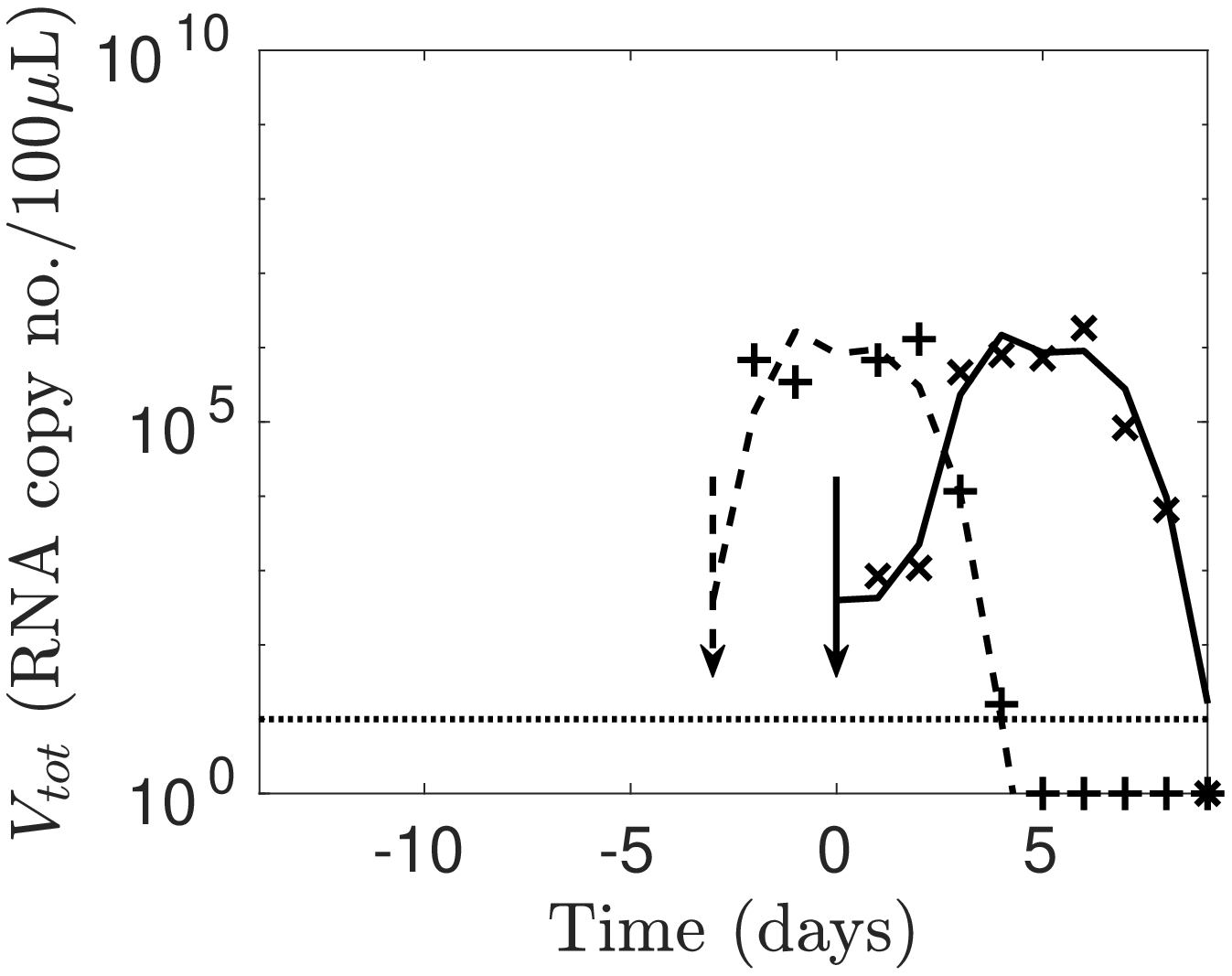}
\caption{3-day interval}
\end{subfigure}

\begin{subfigure}[t]{.32\textwidth}
\captionsetup{justification=centering}
\includegraphics[width =\textwidth]{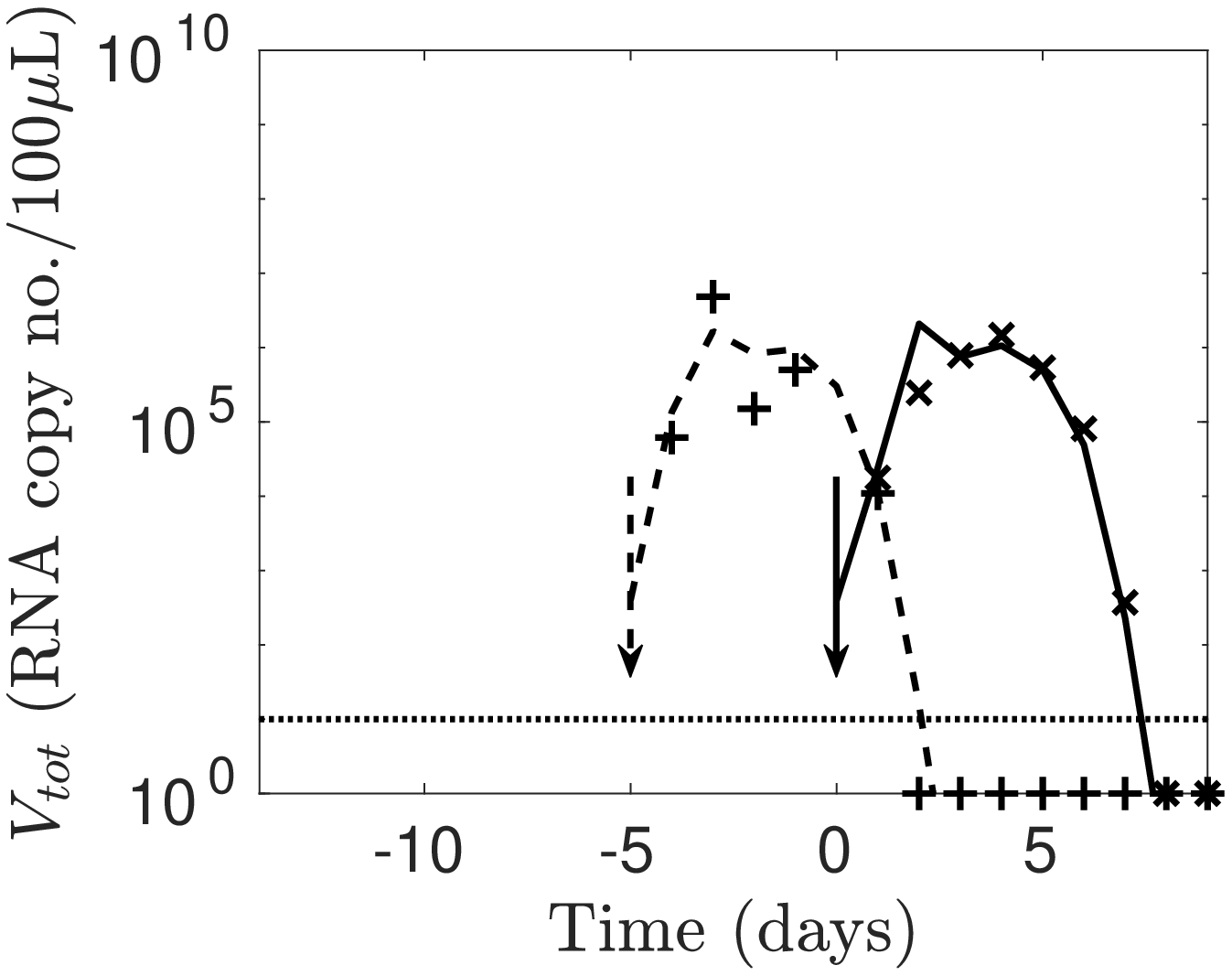}
\caption{5-day interval}
\end{subfigure}
\begin{subfigure}[t]{.32\textwidth}
\captionsetup{justification=centering}
\includegraphics[width =\textwidth]{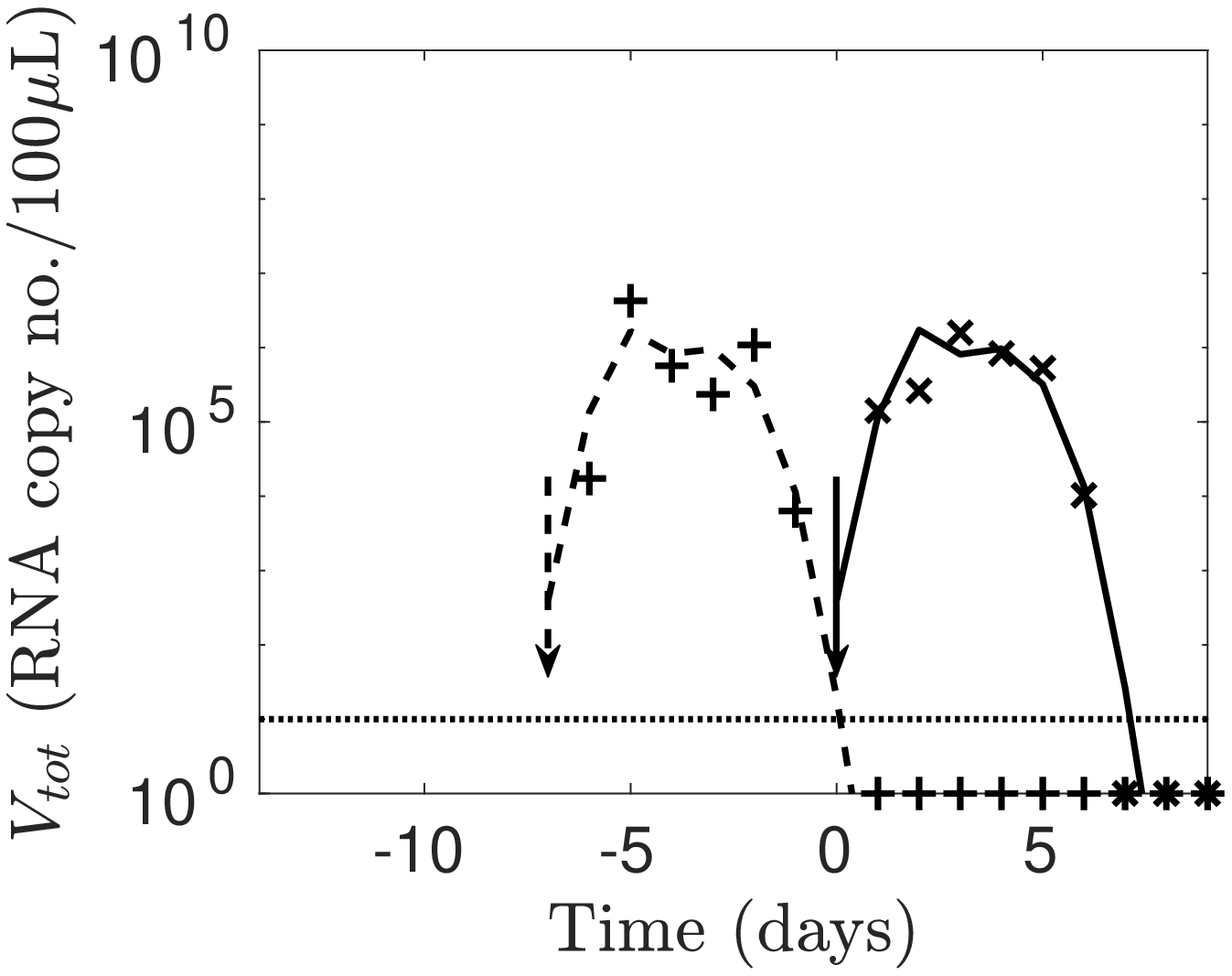}
\caption{7-day interval}
\end{subfigure}
\begin{subfigure}[t]{.32\textwidth}
\captionsetup{justification=centering}
\includegraphics[width =\textwidth]{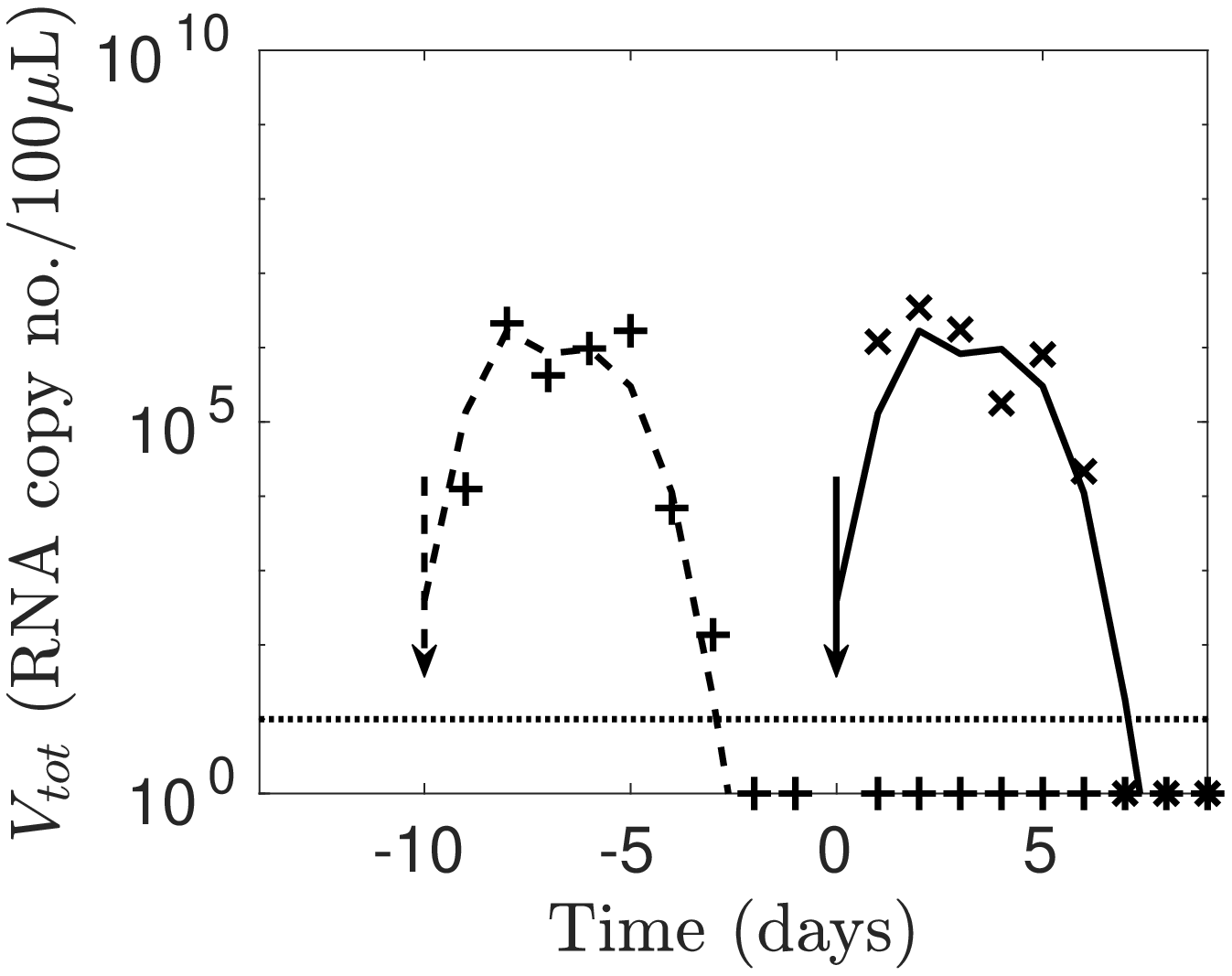}
\caption{10-day interval}
\end{subfigure}

\begin{subfigure}[t]{.32\textwidth}
\captionsetup{justification=centering}
\includegraphics[width =\textwidth]{figs_main/data_14.eps}
\caption{14-day interval}
\end{subfigure}
\paragraph*{S1 Fig}
\label{fig:all_data}
{\bf The full set of synthetic data.}
(a) The line shows the simulated `true' viral load for a single infection, with the arrow showing the time of exposure.
The simulated observed viral loads for the thirteen single infection ferrets are shown as crosses.
The horizontal line indicates the observation threshold (10 RNA copy no./100$\mu$L); observations below this threshold are plotted below this line.
Values below the observation threshold were treated as censored.
(b--g) For sequential infections with the labelled inter-exposure interval, the dashed and dotted lines show the simulated `true' viral load for a primary and challenge infection respectively; the arrows show the times of the primary and challenge exposures.
The simulated observed viral load is shown as crosses.
The sequential infection dataset consists of the viral load for the six sequential infection ferrets and one single infection ferret; the single infection dataset consists of the viral load for the thirteen single infection ferrets.
\end{figure}

Fig~\ref{fig:data} shows a subset of the synthetic data.
For a single infection, the viral load trajectory can be split into exponential growth, plateau and decay phases.
For short inter-exposure intervals (1--5 days), infection with the challenge virus was delayed; for long inter-exposure intervals (7--14 days), infection with the challenge virus was unaffected.
These features of the synthetic data match the qualitative results of Laurie \textit{et al.}~\cite{Laurie2015} for infection with influenza A followed by influenza B, or vice versa.
The parameter values were chosen such that the delay was due to innate immunity.
This choice was made because experimentally, innate immune markers such as type I interferon were observed to be elevated 1--5 days after a primary infection \cite{Carolan2016}, and our previous mathematical model incorporating the innate immune response made predictions consistent with the observed temporary immunity \cite{Cao2015}.
The full set of synthetic data is provided in \nameref{fig:all_data}.

\subsection*{Verification of the fitting procedure}

In this section, we first verify that we could recover the simulated `true' viral load by fitting our model to the data.
In the next section, we will sample from the joint posterior distributions thus obtained to extract the contribution of each immune component.

\begin{figure}[!h]
\begin{adjustwidth}{-2.25in}{0in}
\centering
\begin{subfigure}[c]{.32\textwidth}
\captionsetup{justification=centering}
\includegraphics[width =\textwidth]{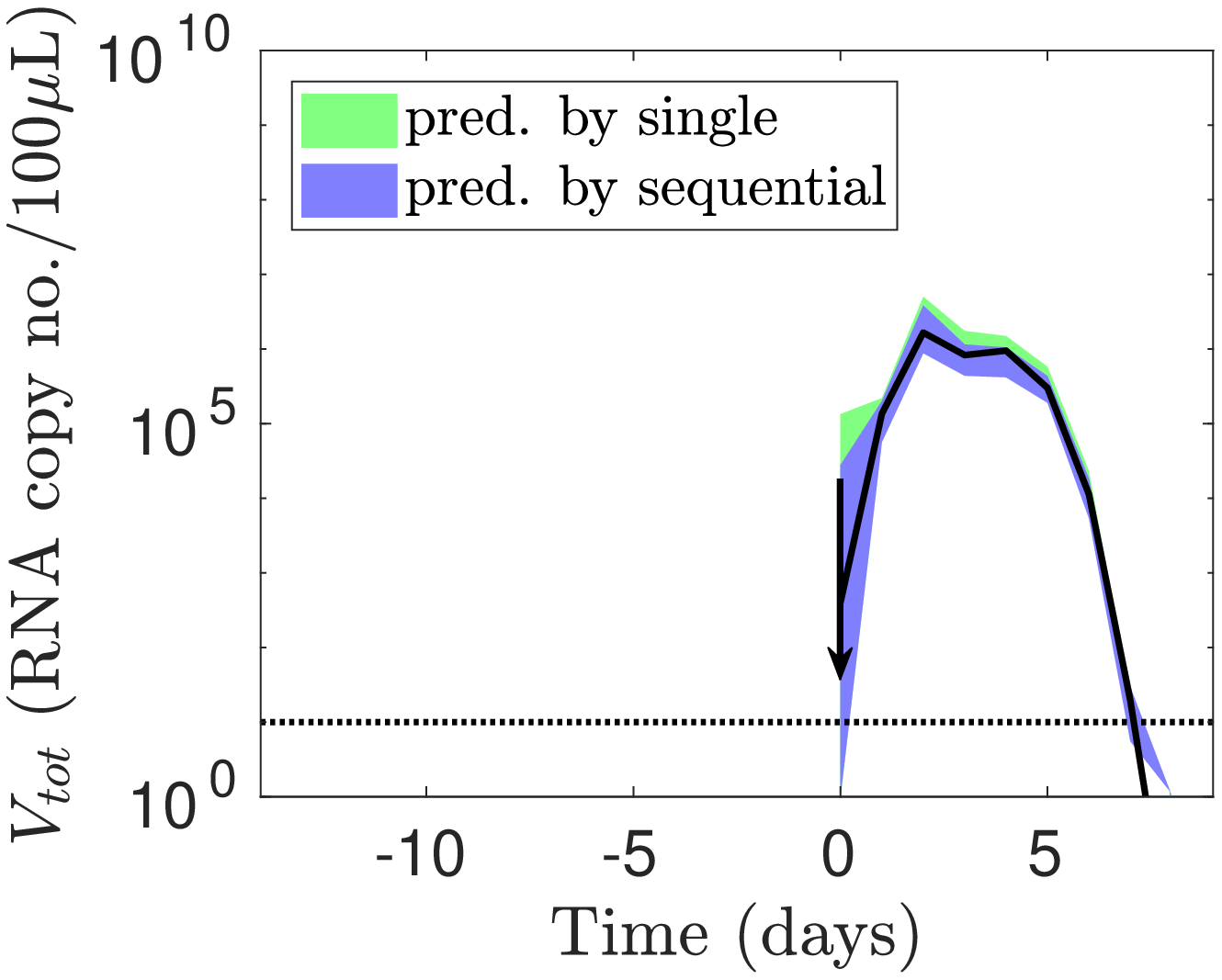}
\caption{single infection}
\end{subfigure}
\begin{subfigure}[c]{.32\textwidth}
\captionsetup{justification=centering}
\includegraphics[width =\textwidth]{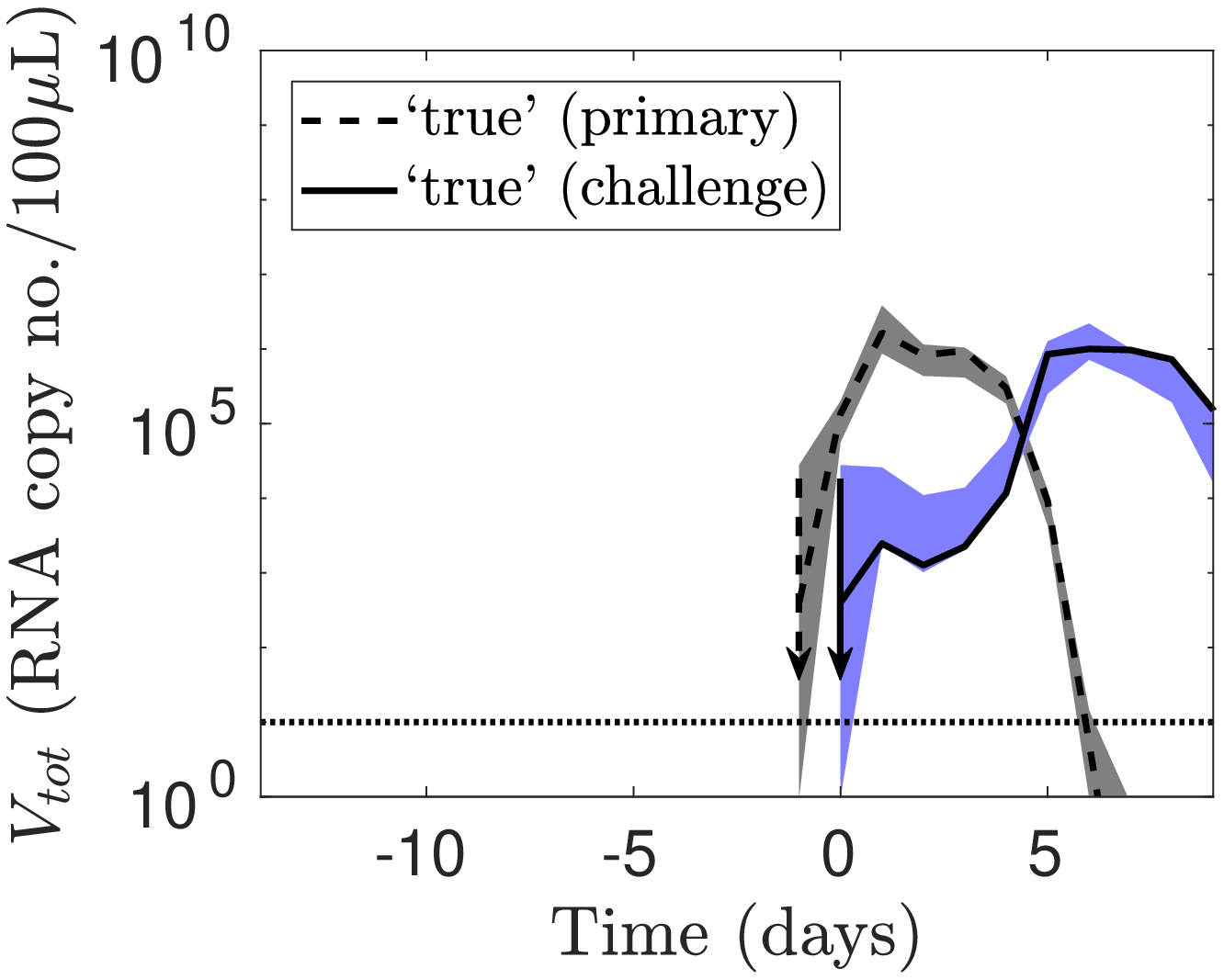}
\caption{1-day interval}
\end{subfigure}
\begin{subfigure}[c]{.32\textwidth}
\captionsetup{justification=centering}
\includegraphics[width =\textwidth]{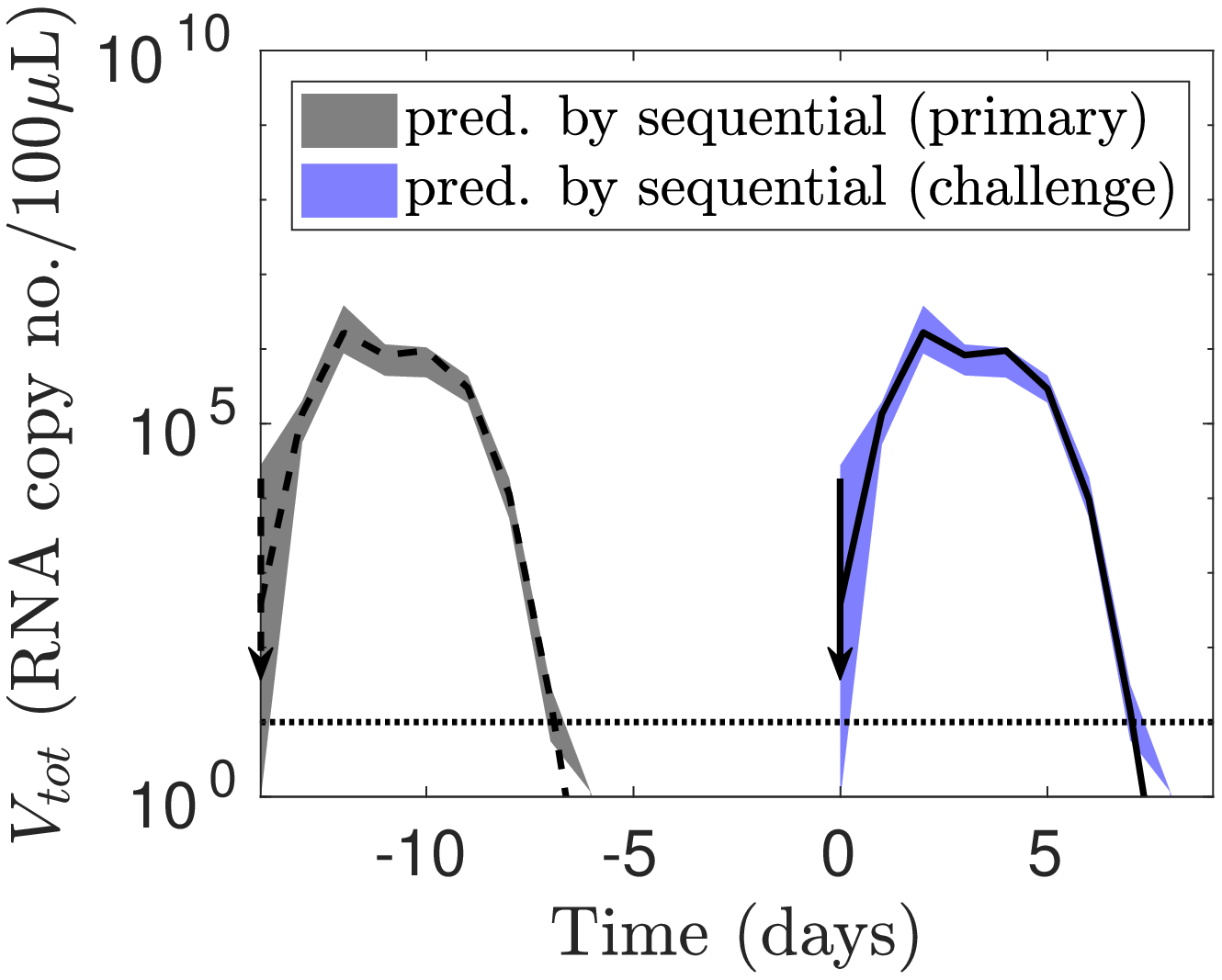}
\caption{14-day interval}
\end{subfigure}
\caption{{\bf Verification that the fitting procedure recovered the viral load.}
(a) For a single infection, the blue and green areas are the 95\% credible intervals for the viral load (in the absence of noise), as predicted by the models fitted to the sequential infection and single infection data respectively.
(b--c) For sequential infections with the labelled inter-exposure interval, the grey and blue areas show the 95\% credible intervals for the primary and challenge viral load respectively, predicted by the model fitted to sequential infection data.
The other elements of the figure are identical to Fig~\ref{fig:data}: the dashed and dotted lines show the simulated `true' viral load for a primary and challenge infection respectively; the arrows show the times of the primary and challenge exposures; and the horizontal line indicates the observation threshold.}
\label{fig:model_fit}
\end{adjustwidth}
\end{figure}

Fig~\ref{fig:model_fit}a shows that the simulated `true' viral load was recovered accurately when fitting the model to either sequential infection or single infection data.
The blue and green (overlapping) areas are 95\% credible intervals predicted by the models fitted to the sequential infection and single infection data respectively.
Both shaded areas included the simulated `true' viral load, shown as the dotted line.
This consistency indicates that the fitting procedure accurately recovered the viral load.

Figs~\ref{fig:model_fit}b--c confirm that fitting to sequential infection data accurately recovered the viral load for different inter-exposure intervals.
The grey and blue areas show the 95\% credible intervals for the primary and challenge viral load respectively.



\subsection*{Comparing the immunological information in each dataset}

Next, we compared the behaviour of the fitted models to the behaviour of the `true' parameters, to determine the information in each dataset on

\begin{itemize}
	\item the effect of each immune component in controlling a single infection;
	\item cross-protection between strains; and
	\item each immune component's contribution to cross-protection.
\end{itemize}

\subsubsection*{The effect of each immune component in controlling a single infection}

\begin{figure}[h!]
\begin{adjustwidth}{-2.25in}{0in}
\centering
\begin{subfigure}[t]{.32\textwidth}
\captionsetup{justification=centering}
\includegraphics[width =\textwidth]{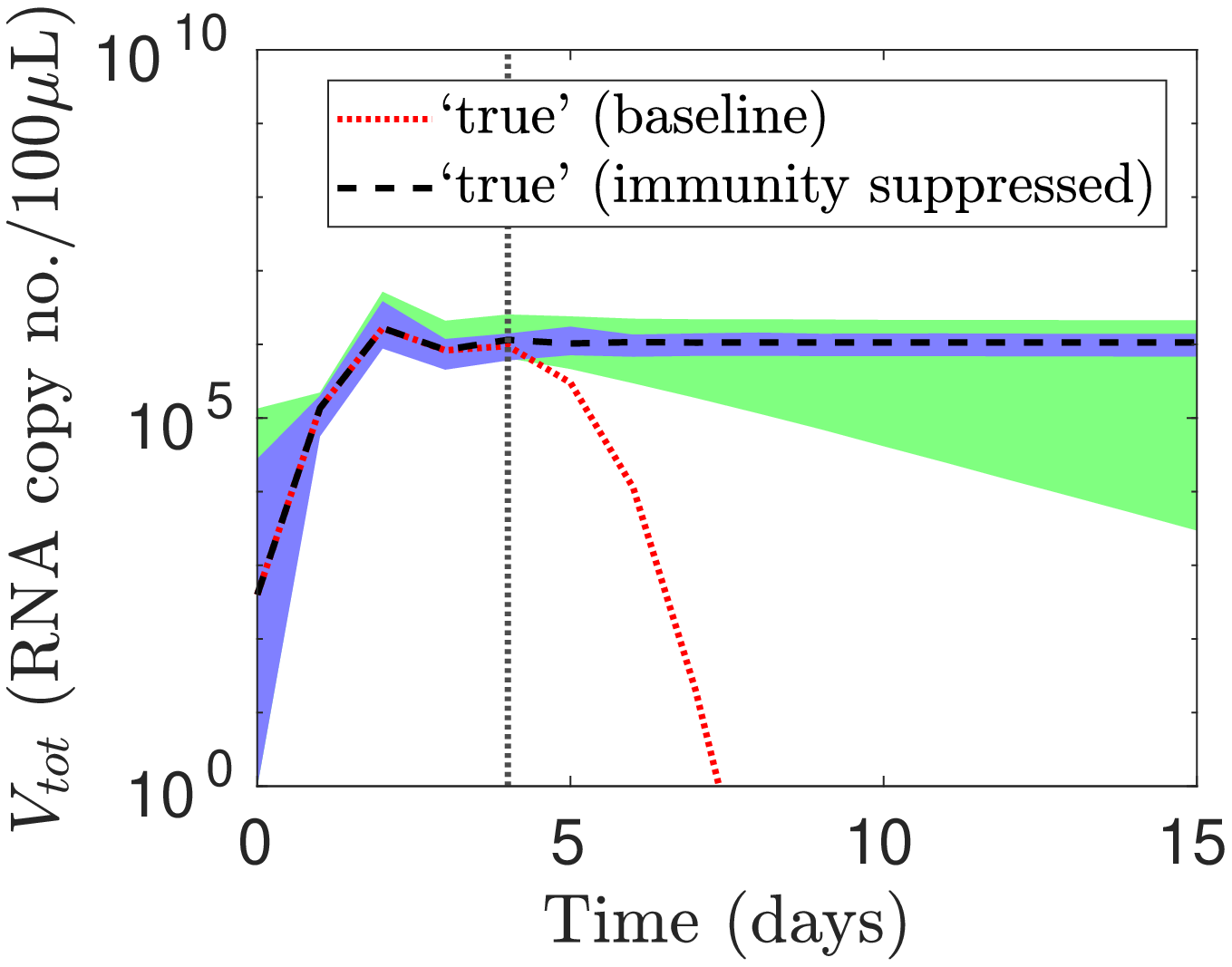}
\caption{adaptive}
\end{subfigure}
\begin{subfigure}[t]{.32\textwidth}
\captionsetup{justification=centering}
\includegraphics[width =\textwidth]{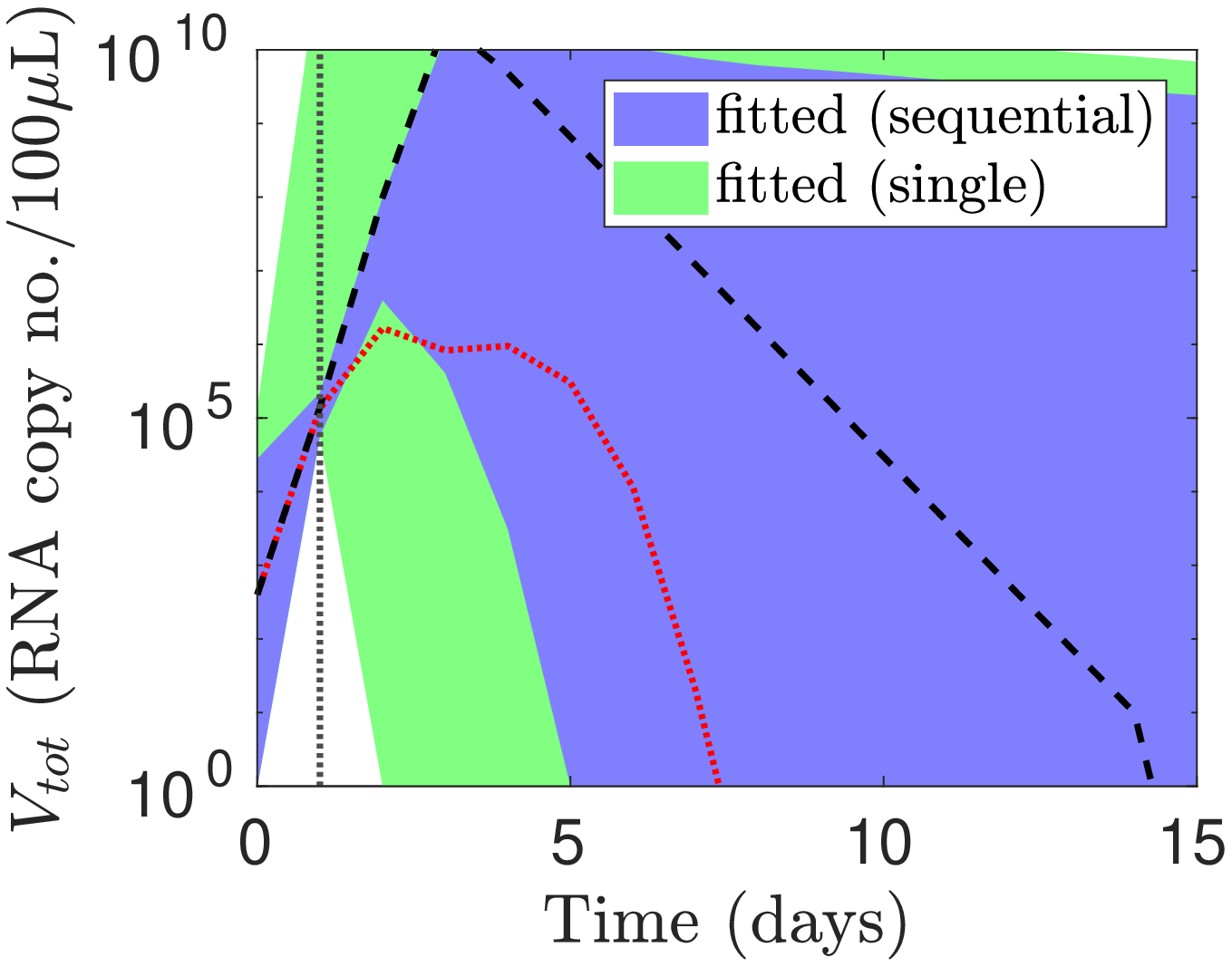}
\caption{innate and adaptive}
\end{subfigure}
\begin{subfigure}[t]{.32\textwidth}
\captionsetup{justification=centering}
\includegraphics[width =\textwidth]{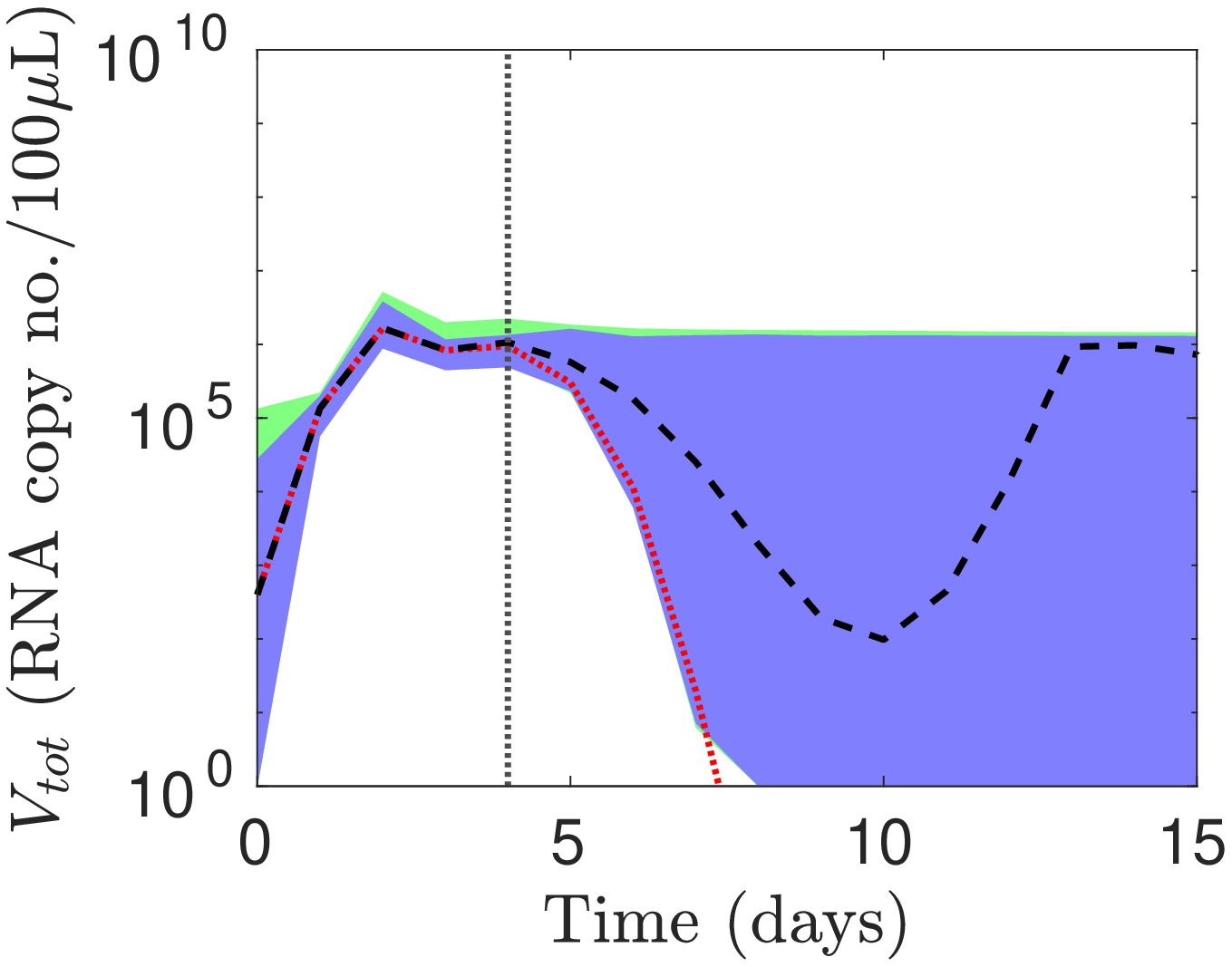}
\caption{humoral adaptive}
\end{subfigure}
\begin{subfigure}[t]{.32\textwidth}
\captionsetup{justification=centering}
\includegraphics[width =\textwidth]{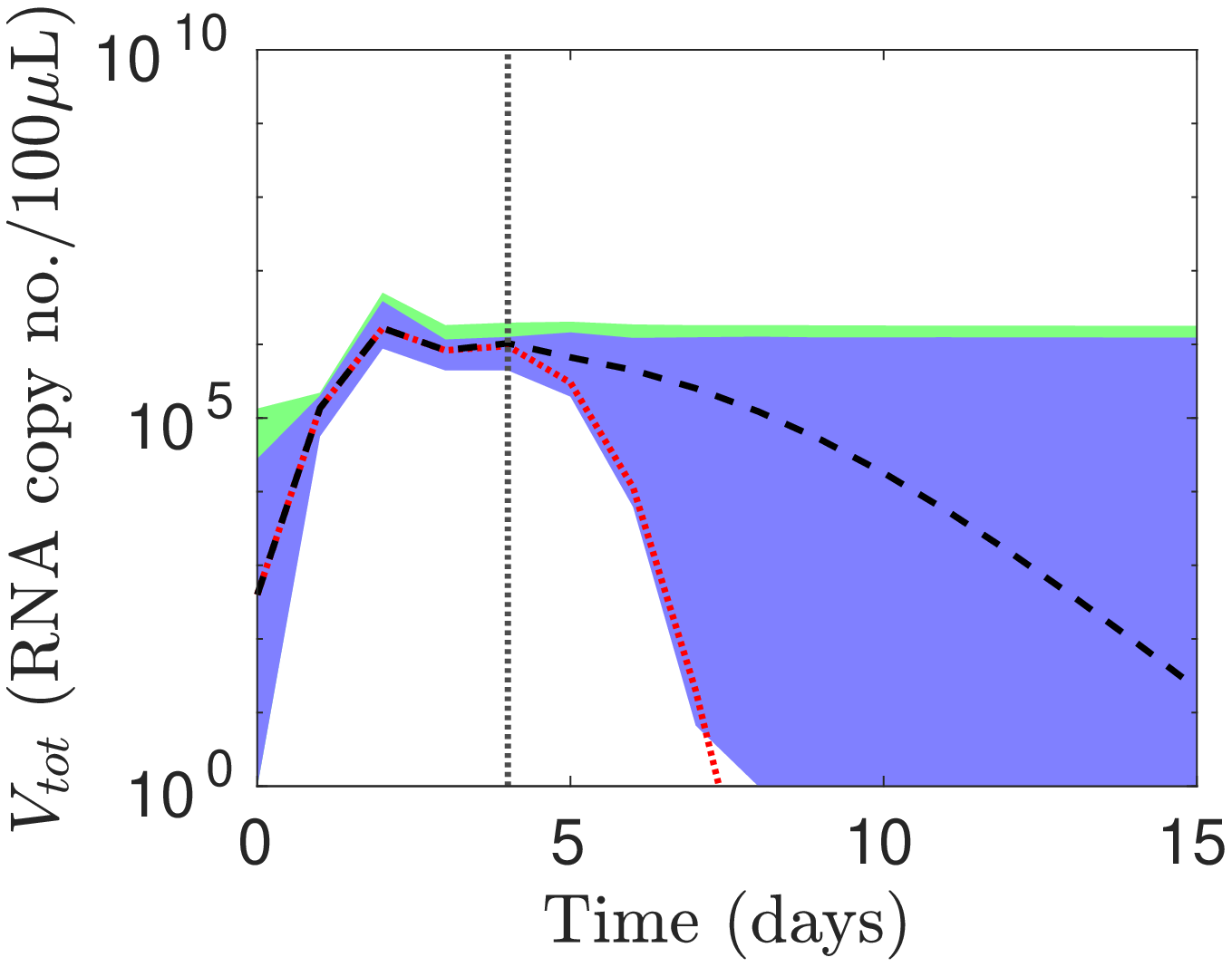}
\caption{cellular adaptive}
\end{subfigure}
\caption{{\bf Predicting the viral load for a single infection when various immune components were absent.}  
The vertical lines indicate, for the `true' parameter values, the times at which the immune components labelled under each panel took effect.
These times were determined by when the viral load for the baseline model (red dotted line) deviated from the viral load when the immune components were absent (black dashed line).
These times were recovered using sequential infection data in all of the panels (95\% prediction intervals for the viral load in blue), while the timing of adaptive immunity in (a) was recovered using single infection data (intervals in green).
In addition, the viral load when adaptive immunity was suppressed was accurately predicted using sequential infection data (a).
However, the viral load was not accurately predicted using either dataset in the remaining scenarios (b--d).
Prediction intervals were constructed without measurement noise.}
\label{fig:knockout}
\end{adjustwidth}
\end{figure}

In Fig~\ref{fig:knockout}, we removed adaptive immunity, both innate and adaptive immunity, humoral adaptive immunity, or cellular adaptive immunity from the model.
We then compared predictions of the viral load for a single infection by the models fitted to the two datasets.

First, we showed how the viral load trajectory for the `true' parameters changed when adaptive immunity was suppressed.
We defined the `baseline' as the viral load when all immune components were present (red dotted lines in Fig~\ref{fig:knockout}, which are the same as the black lines in Figs~\ref{fig:data}a and ~\ref{fig:model_fit}a).
Suppressing adaptive immunity prevented resolution of the infection (black dashed line in Fig~\ref{fig:knockout}a), which was consistent with findings of a previous experiment~\citep{Kris1988}.
The viral load deviated from the baseline trajectory at 4 days post-exposure (vertical line), indicating that this was the time at which adaptive immunity took effect.

We then asked whether the models fitted to the two datasets predicted this change.
Chronic infection in the absence of adaptive immunity was only predicted using sequential infection data (Fig~\ref{fig:knockout}a).
Single infection data did not enable consistent prediction of this outcome, as indicated by the broadening prediction interval.
However, both datasets enabled recovery of the time at which the viral loads in the presence and absence of adaptive immunity deviated (the vertical line in Fig~\ref{fig:knockout}a).
Hence, the timing of adaptive immunity was accurately estimated using either dataset.

In Figs~\ref{fig:knockout}b--d, we repeated this procedure, suppressing (b) all immunity; (c) humoral adaptive immunity; or (d) cellular adaptive immunity.
When both innate and adaptive immunity were suppressed, the peak viral load increased, and resolution of the infection was delayed.
These changes were consistent with a previous experiment where innate immunity was suppressed~\cite{Seo2002}.

Fig~\ref{fig:knockout}b shows that sequential infection data enabled accurate inference of when the viral loads in the presence and absence of immunity deviated, hence recovering the timing of innate immunity.
By contrast, the model fitted to single infection data predicted that the viral loads could deviate much earlier.
Neither model accurately predicted how the infection resolved in the absence of the immune response; however, the prediction intervals for the model fitted to sequential infection data were tighter, and the peak viral load was consistently predicted to be higher than for the baseline model.

In the absence of the immune response, the infection resolves due to target cell depletion only.
The lack of predictive ability indicates that both datasets lack information on how target cells would hypothetically become depleted, and how this depletion would affect the viral load, in the absence of the immune response.
One is thus cautioned against using parameter values from a model fitted to data in immunocompetent hosts to make predictions in situations where target cells may become severely depleted, such as if individuals are immunocompromised.

Figs~\ref{fig:knockout}c--d show that neither dataset enabled prediction of how the viral load changed when (c) the humoral adaptive immune response or (d) the cellular adaptive immune response was removed.
This implies that sequential infection data (of the type reported in \Laurie{}) cannot be used to distinguish the contributions of antibodies and cellular adaptive immunity to resolution of infection.
In detail, the `true' parameters predicted that when humoral adaptive immunity was disabled, the viral load rebounded instead of continuing to decrease (black dashed line in Fig~\ref{fig:knockout}c).
When cellular adaptive immunity was disabled, resolution of the infection was delayed (black dashed line in Fig~\ref{fig:knockout}d).
The fitted model's predictions ranged from no delay to a chronic infection.

\subsubsection*{Cross-protection between strains}

Given the above mixed results, we then tested whether sequential infection data accurately captured the timing and extent of cross-protection, by simulating the viral load for inter-exposure intervals other than those where data was provided.
We selected new inter-exposure intervals of 2 and 20~days; the former lay between inter-exposure intervals included in the original data (1, 3, 5, 7, 10 and 14 days), while the latter lay outside this range.
Then, using the models fitted to the original data (that is, not re-fitting to the new data), we predicted the challenge viral load for these new inter-exposure intervals.
Because a primary infection could greatly affect a challenge infection, but not vice versa, we focused on the behaviour of the challenge infection.

\begin{figure}[!h]
\begin{adjustwidth}{-2.25in}{0in}
\centering
\begin{subfigure}[c]{.32\textwidth}
\captionsetup{justification=centering}
\includegraphics[width =\textwidth]{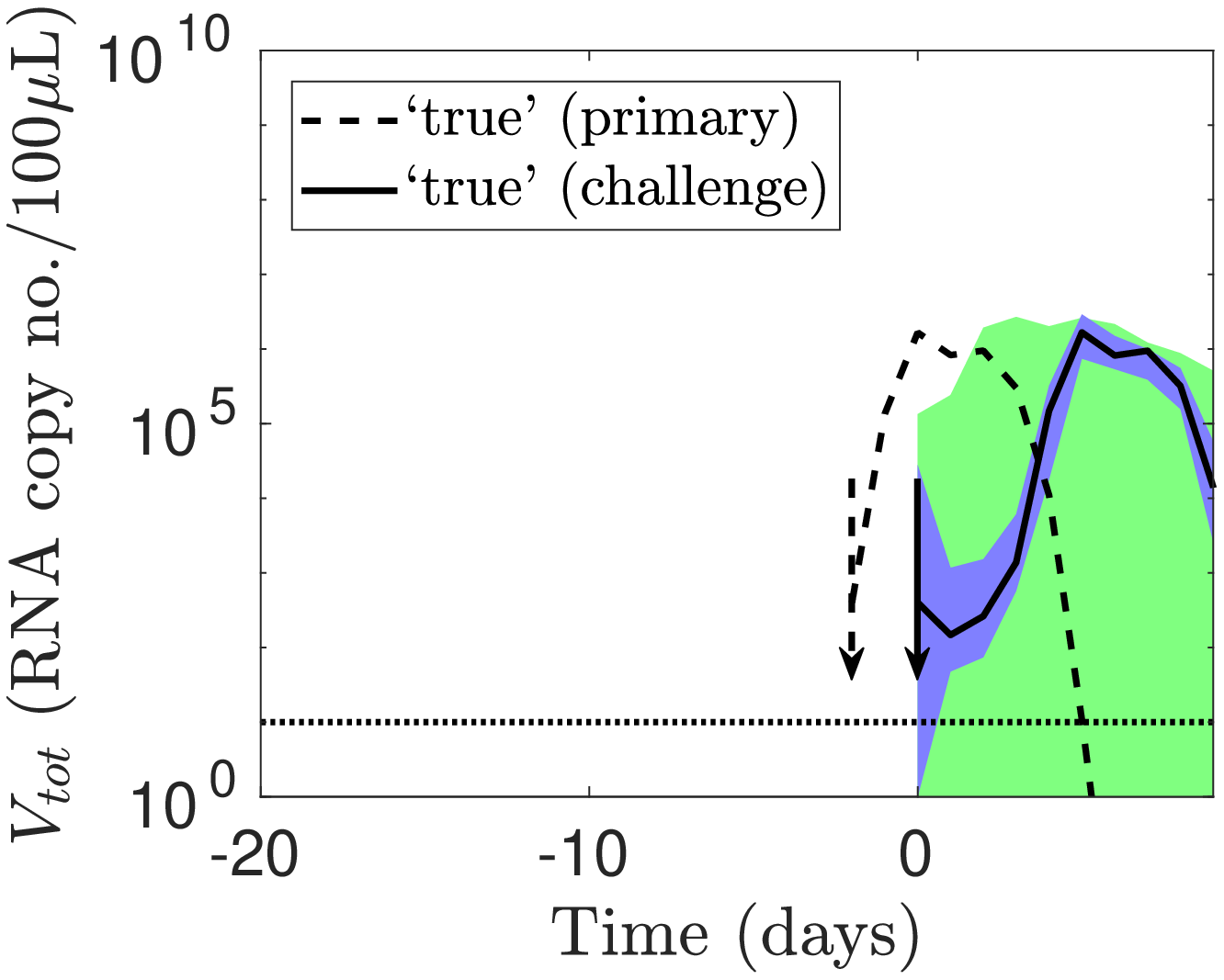}
\caption{2-day interval}
\end{subfigure}
\begin{subfigure}[c]{.32\textwidth}
\captionsetup{justification=centering}
\includegraphics[width =\textwidth]{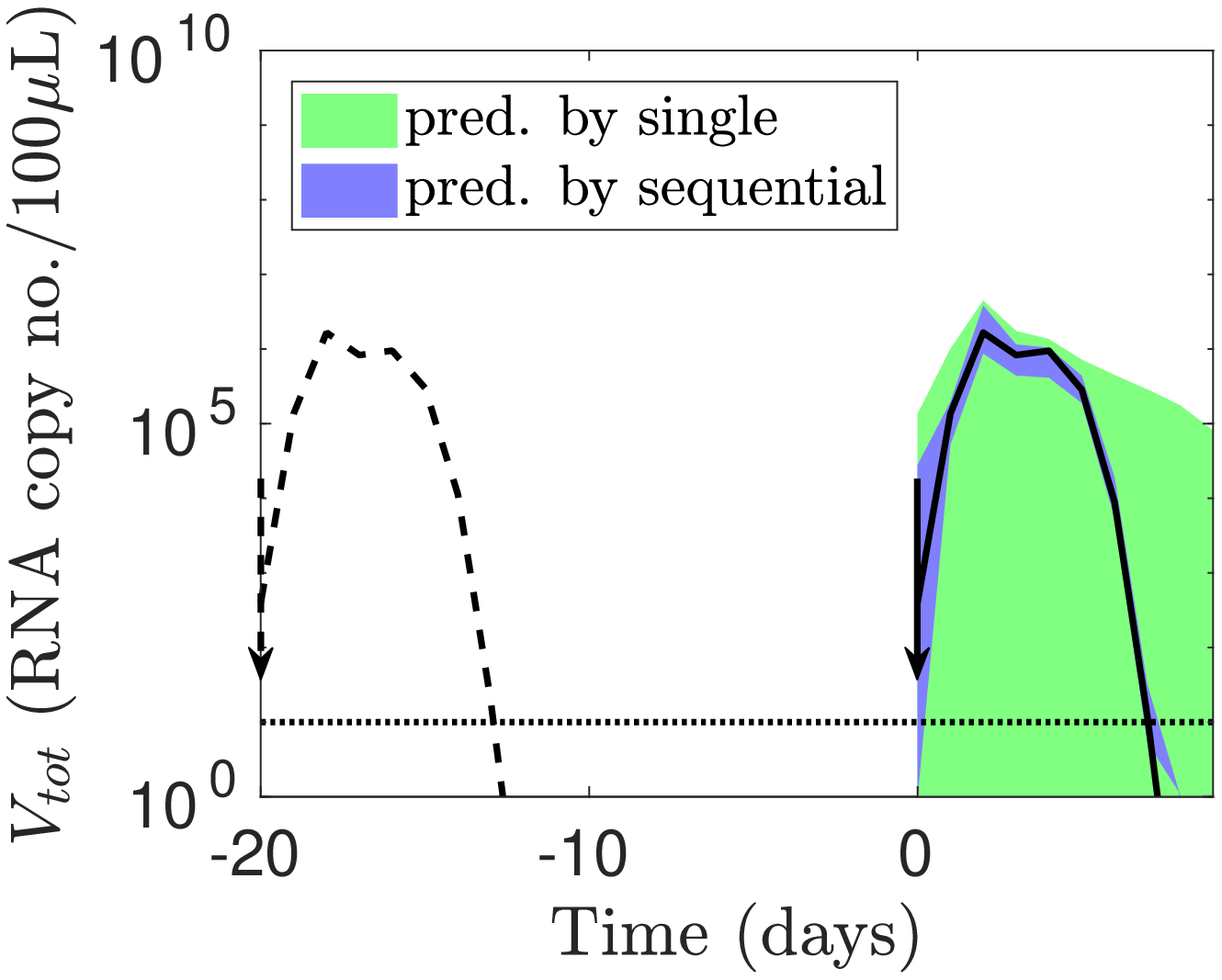}
\caption{20-day interval}
\end{subfigure}
\caption{{\bf Predicting the outcomes of further sequential infection experiments.}
Sequential infection data, but not single infection data, enabled prediction of further sequential infection experiment outcomes.
The lines show the simulated `true' viral loads for inter-exposure intervals of 2 and 20 days.  The shaded areas show the 95\% prediction intervals for the challenge viral load.}
\label{fig:predict_other_intervals}
\end{adjustwidth}
\end{figure}

Fig~\ref{fig:predict_other_intervals} shows that predictions by the model fitted to sequential infection data (blue areas) were accurate.
By contrast, predictions using single infection data (green) did not agree well with the `true' viral load.
Note that to predict cross-protection using single infection data, we used the model assumptions that innate immunity was completely non-specific and antibodies were completely strain-specific, and considered the optimistic scenario where we had independent, perfect information about the proportion of cellular adaptive immunity that was cross-reactive (details in the \methods{} section).
Even then, single infection data did not accurately capture cross-protection.

\subsubsection*{Each immune component's contribution to cross-protection}

Having shown that the sequential infection data captures the timing and extent of cross-protection between strains, we then asked whether such cross-protection could be attributed to the `correct' mechanisms (the same mechanisms as given by the `true' parameters).
These mechanisms are

\begin{itemize}
	\item target cell depletion due to the infection and subsequent death of cells;
	\item innate immunity; and 
	\item cellular adaptive immunity.
\end{itemize}
In our model, antibodies are strain-specific and thus do not contribute to cross-protection.

\begin{figure}[h!]
\begin{adjustwidth}{-2.25in}{0in}
\centering
\begin{subfigure}[c]{.32\textwidth}
\captionsetup{justification=centering}
\includegraphics[width =\textwidth]{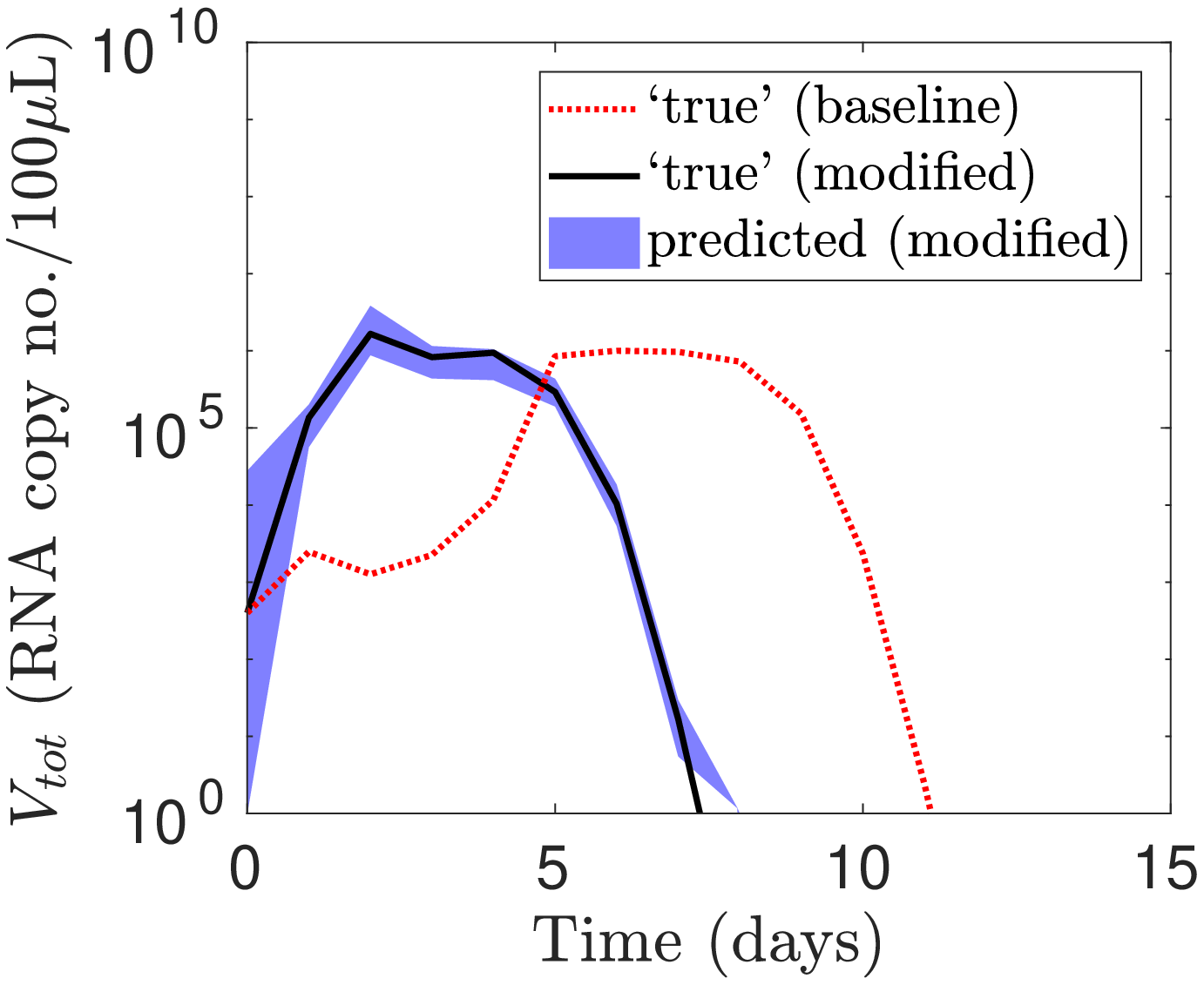}
\caption{Model XC}
\end{subfigure}
\begin{subfigure}[c]{.32\textwidth}
\captionsetup{justification=centering}
\includegraphics[width =\textwidth]{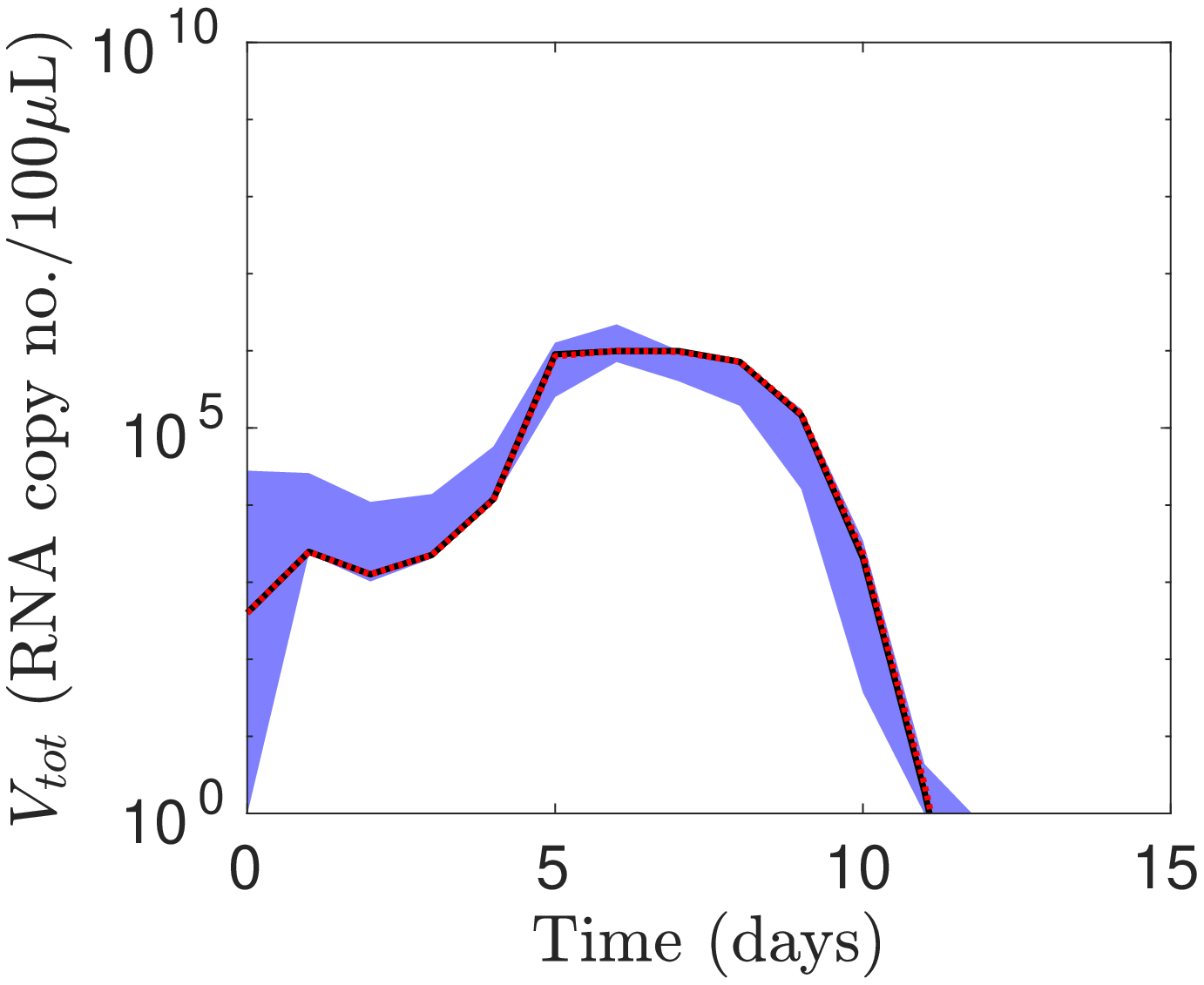}
\caption{Model XIT}
\end{subfigure}
\begin{subfigure}[c]{.32\textwidth}
\captionsetup{justification=centering}
\includegraphics[width =\textwidth]{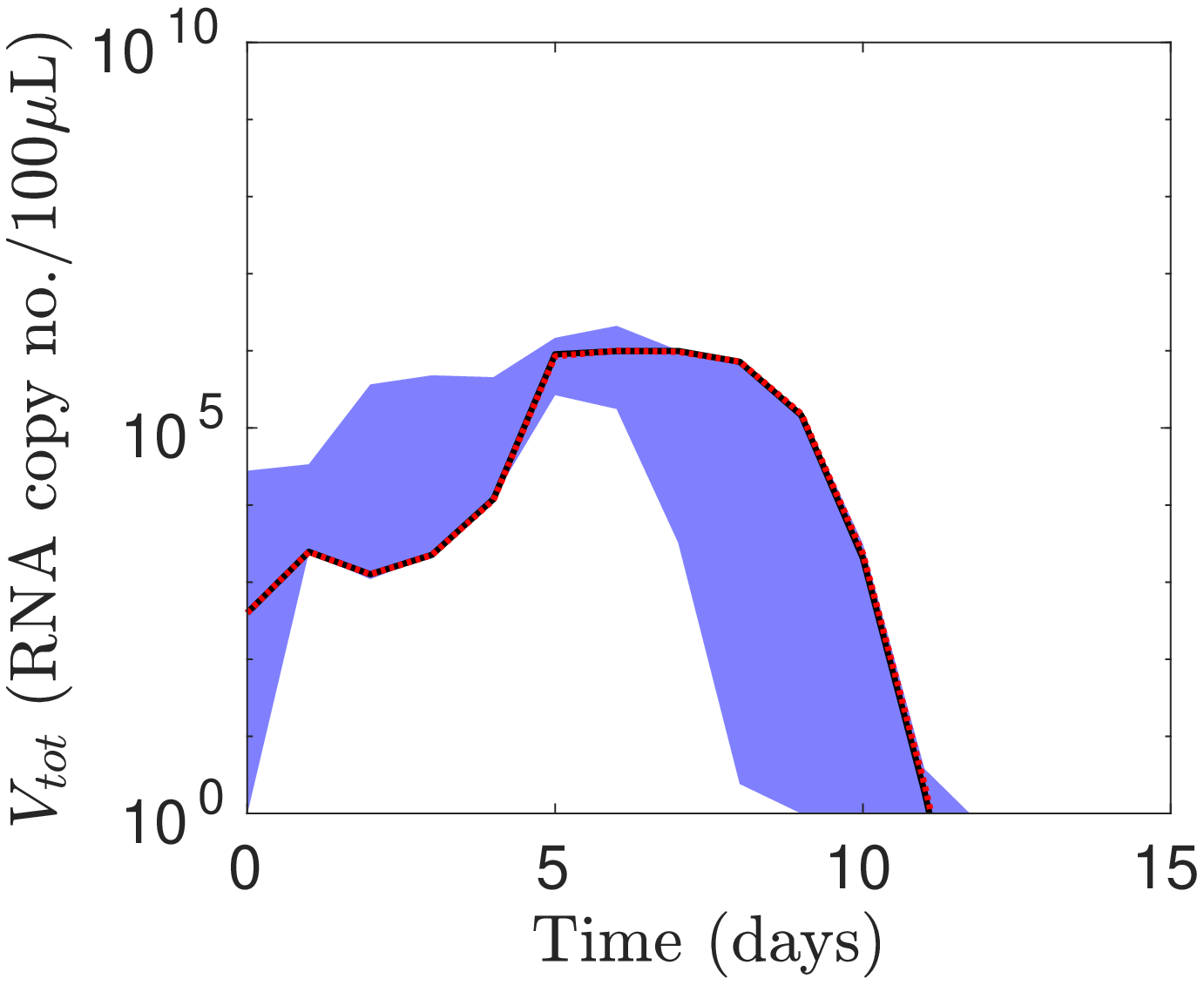}
\caption{Model XI}
\end{subfigure}
\caption{{\bf Predictions of the challenge viral load for a one-day inter-exposure interval when the mechanisms mediating cross-protection were restricted.}  
The red dotted lines show the challenge viral load for the `true' parameter values when the mechanisms mediating cross-protection were restricted.
The black solid lines show the viral load for the baseline model.
Comparing the two sets of lines revealed that innate immunity mediated cross-protection, whereas cellular adaptive immunity did little to mediate cross-protection.
The model fitted to sequential infection data accurately predicted the challenge outcomes for models XC and XIT, but not model XI (95\% prediction intervals shown).
It thus correctly attributed cross-protection to target cell depletion and/or innate immunity, but could not definitively distinguish between the two.
The viral load for the primary infection is not shown, to improve clarity of the figure.}
\label{fig:xreact_components}
\end{adjustwidth}
\end{figure}

Before analysing the behaviour of the fitted models, we quantified how each immune component contributed to cross-protection for the `true' parameters.
In Fig~\ref{fig:xreact_components}, for a one-day inter-exposure interval, we plotted in red the challenge viral load for the baseline model (the original model fitted to the data, where all three of the above immune components can mediate cross-protection).
We observed that the challenge infection was delayed relative to a primary infection.
We then modified the baseline model such that only a subset of immune components mediates cross-protection, as detailed in the \methods{} section.
We used the modified model to predict the viral load (in black), and compared it with the baseline viral load.
(The blue areas will be discussed shortly.)

For example, in Fig~\ref{fig:xreact_components}a, we modified the baseline model such that only cellular adaptive immunity, and not target cell depletion or innate immunity, can mediate cross-protection.
We denoted this modified model `model XC'.
Unlike the baseline model (red dotted line), the challenge viral load for model XC was not delayed (black solid line); in fact, it closely resembled that for a single infection.
Comparing the two simulations led to the conclusion that cellular adaptive immunity did not play a major part in cross-protection.

We then modified the baseline model such that only target cell depletion and/or innate immunity, but not cellular adaptive immunity, can mediate cross-protection.
We denoted this model `model XIT'.
The challenge viral loads according to model XIT and the baseline model were very similar (overlapping lines in Fig~\ref{fig:xreact_components}b).
Hence, for the `true' parameters, cross-protection was mediated by innate immunity and/or target cell depletion.

To distinguish between these two mechanisms, we constructed model XI, where only innate immunity, and not target cell depletion or cellular adaptive immunity, can mediate cross-protection.
Once again, the challenge viral load was very similar to the baseline model.
We concluded that the cross-protection was largely mediated by innate immunity.

Having demonstrated the utility of the modified models, we returned to the original question of whether sequential infection data could be used to distinguish between mechanisms for cross-protection.
We sampled parameter sets from the joint posterior distributions obtained by fitting the baseline model to sequential infection data, and used them as inputs for models XC, XIT and XI respectively, to generate the blue areas in Fig~\ref{fig:xreact_components}.
If the fitted parameters and the `true' parameters predict the same infection outcomes under the modified models, then the fitted model attributes cross-protection to the `correct' mechanisms.

\begin{figure}[!h]
\centering
\includegraphics[width =.32\textwidth]{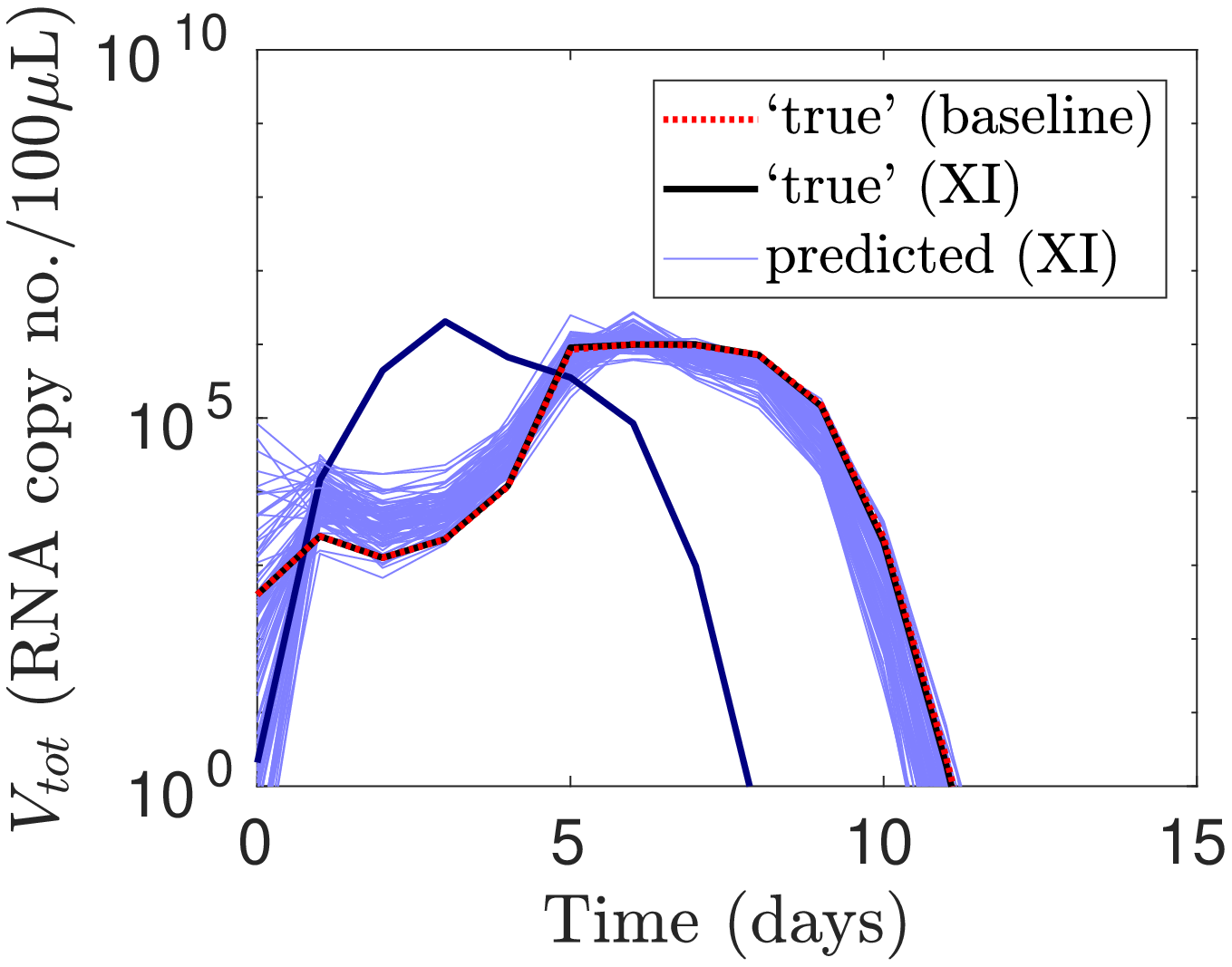}
\paragraph*{S2 Fig}
\label{fig:model_XI_results}
{\bf Trajectories for model XI generated using 100 uniformly sampled parameter sets from the MCMC chains after burn-in, for the model fitted to sequential infection data.}
The darker trajectory incorrectly attributed the delay observed in the baseline model to target cell depletion rather than innate immunity.
\end{figure}

Models XC and XIT made the same predictions using the fitted parameters (shaded area) and the `true' parameters (black line), so sequential infection data enabled us to accurately attribute cross-protection to target cell depletion and/or innate immunity, rather than cellular adaptive immunity.
On the other hand, the fitted parameters did not consistently predict the challenge outcome for model XI (Fig~\ref{fig:xreact_components}b).
Hence, we could not use sequential infection data to consistently rule out the possibility that cross-protection occurred due to target cell depletion rather than innate immunity.
However, only a very small proportion of trajectories sampled from the joint posterior distribution incorrectly attributed the delay to target cell depletion (\nameref{fig:model_XI_results}).

\begin{figure}[!h]
\centering
\begin{subfigure}[c]{.32\textwidth}
\captionsetup{justification=centering}
\includegraphics[width =\textwidth]{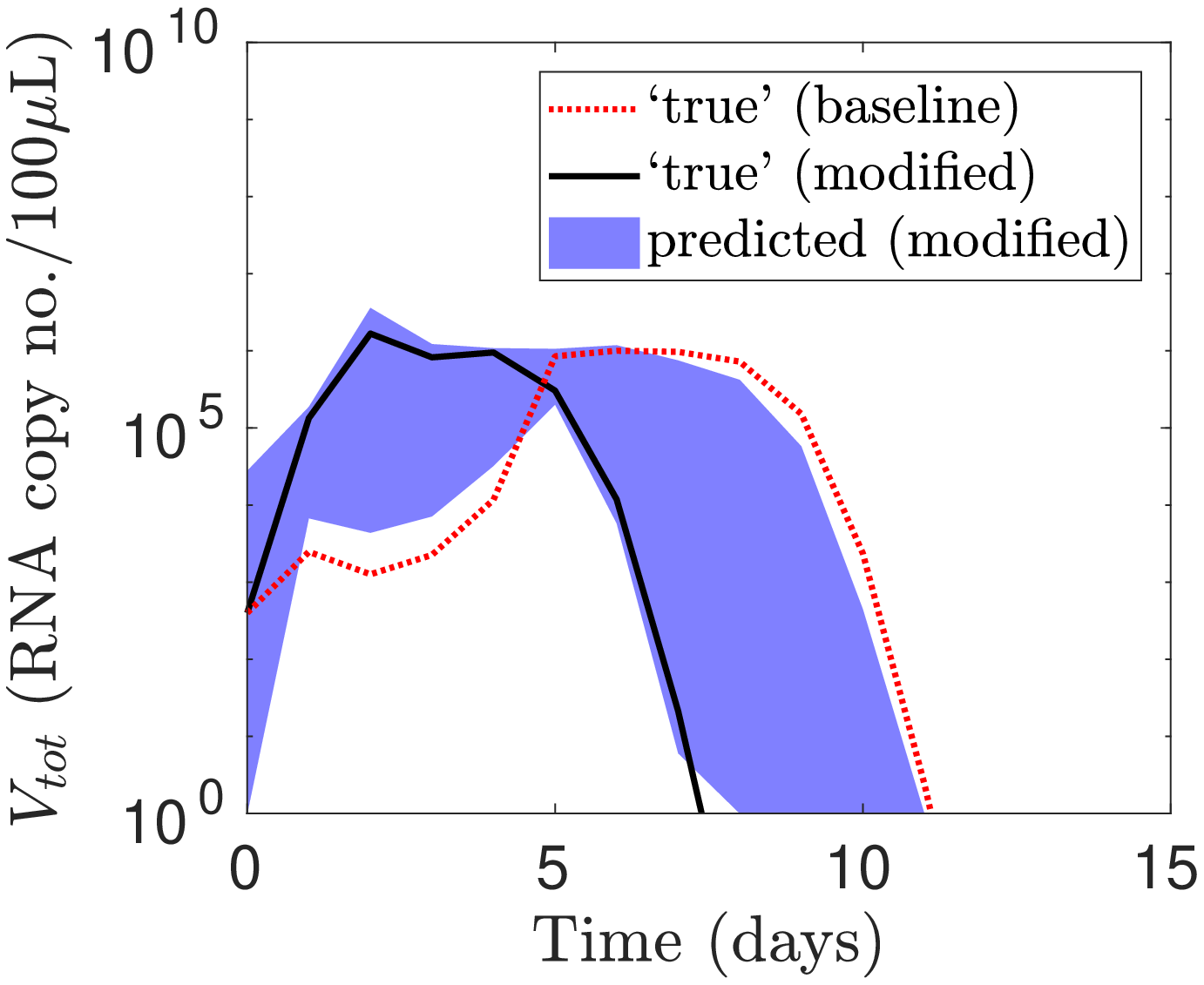}
\caption{Model XI1}
\end{subfigure}
\begin{subfigure}[c]{.32\textwidth}
\captionsetup{justification=centering}
\includegraphics[width =\textwidth]{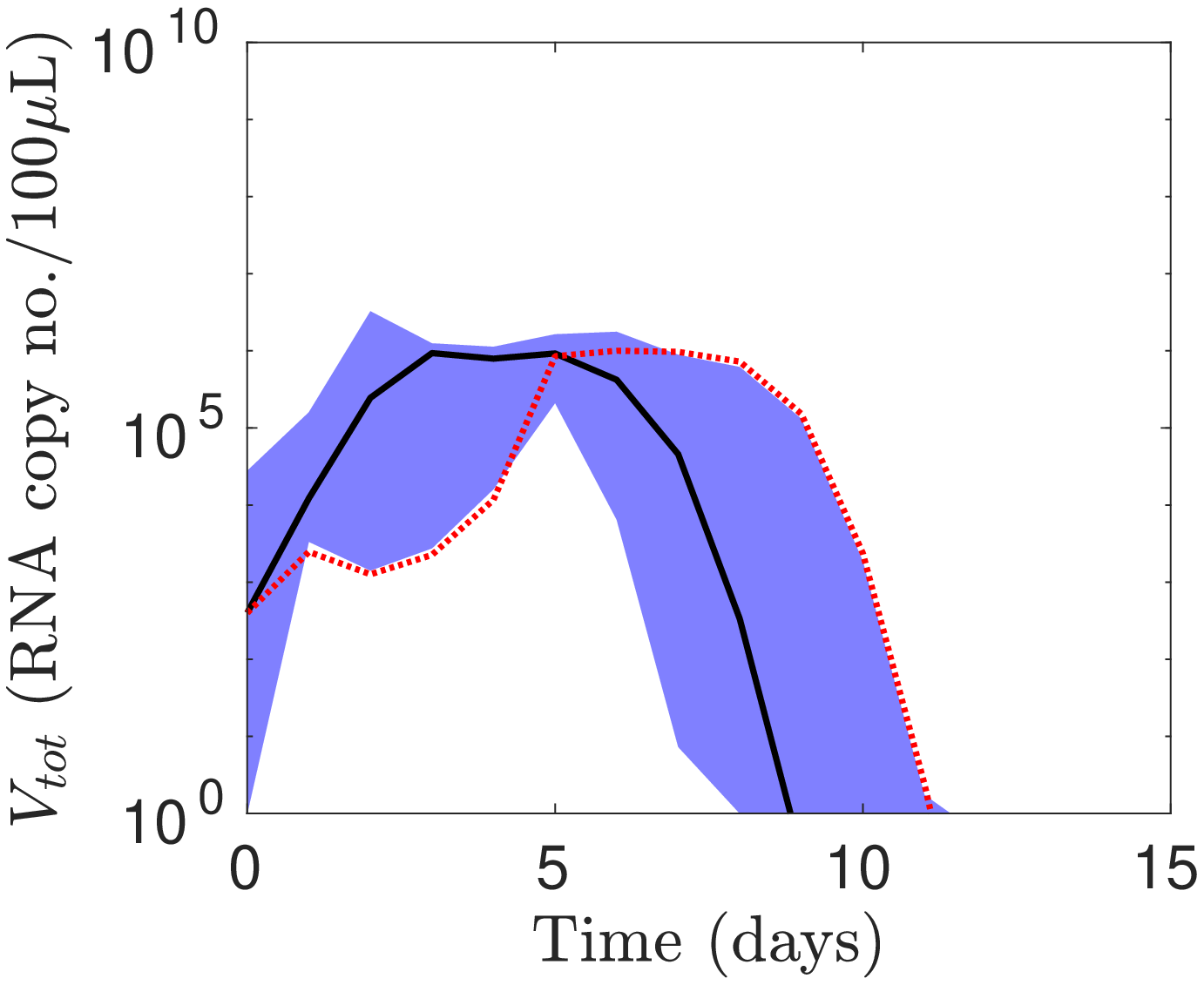}
\caption{Model XI2}
\end{subfigure}
\begin{subfigure}[c]{.32\textwidth}
\captionsetup{justification=centering}
\includegraphics[width =\textwidth]{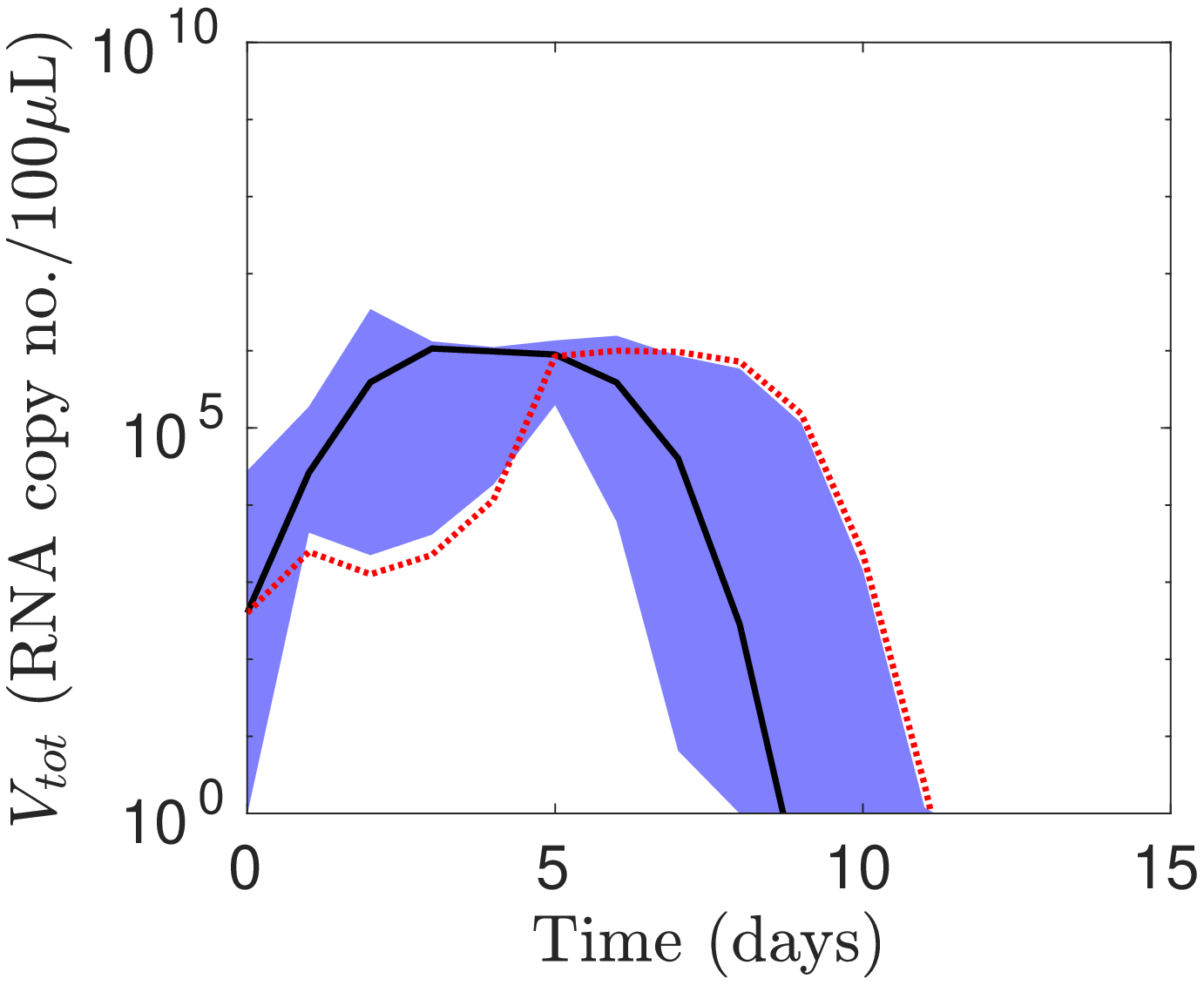}
\caption{Model XI3}
\end{subfigure}
\paragraph*{S3 Fig}
\label{fig:model_XI123_results}
{\bf Sequential infection data did not enable accurate prediction of the challenge viral load for a modified model where only one of the three innate immune mechanisms mediates cross-protection.}
The challenge viral load for the `true' parameter values and a modified model where cross-protection is mediated by only one innate immune mechanism (models XI1--XI3, red line) was compared to the viral load for the baseline model (black line).
At a one-day inter-exposure interval, the delay in the baseline model occurred due to a combination of innate immune mechanisms 2 and 3.
Prediction intervals for the viral load for models XI1--XI3 according to the model fitted to sequential infection data (blue areas) did not accurately recover the viral load according to the `true' parameters.
Hence, the fitted model did not attribute cross-immunity to the correct mechanisms of the innate immune response.
\end{figure}

Similarly, we were unable to disentangle different mechanisms of innate immunity from the sequential infection data alone (\nameref{fig:model_XI123_results} and \nameref{file:eqs}).

For infection with heterologous influenza A strains, rather than the influenza A and B strains discussed thus far, we hypothesise that innate and cellular adaptive immunity contribute to cross-protection at different inter-exposure intervals~\cite{Yan2017}.
\nameref{file:high_xreact} presents the same analysis for this scenario, where we were able to unravel these different contribution using sequential infection data.

In summary, the synthetic sequential infection data enabled accurate inference of the contribution of cellular adaptive immunity to cross-protection, as well as the combined contributions of target cell depletion and innate immunity.
However, using this data alone, we could not conclusively distinguish the contributions of innate immunity and target cell depletion to cross-protection, or distinguish the contributions of different innate immune mechanisms.

\section*{Discussion}

\subsection*{Advantages of sequential infection experiments}

In this study, we have simulated experiments which investigate the interaction of influenza strains through sequential infections, then explored how mathematical models could be applied to the data to gain insight into immune mechanisms.
Our analysis has shown that the sequential infection study design, compared to the single infection study design, provides richer information for inferring the timing and strength of each immune component.

We have identified three main advantages of sequential infection data.
The first advantage is in inferring how each immune component helps to resolve a single infection.
We found that the synthetic sequential infection data captures the timing of innate and adaptive immunity during a single infection, and thus enables accurate prediction of the outcomes of some \textit{in silico} experiments where immune components were removed.
In contrast, we could not consistently infer the timing and strength of innate immunity from single infection data.
Moreover, this type of data contains information only on the timing of adaptive immunity, but not the effects of suppressing adaptive immunity.

The second advantage is that sequential infection data contains more information on the effects of cross-protection.
We were able to use the model fitted to the sequential infection data to precisely predict outcomes of further such experiments using the same strains but different inter-exposure intervals.  
Using the model fitted to the single infection data greatly reduced predictive power.

The third advantage is in inferring the contribution of each immune component to this cross-protection.
The synthetic sequential infection data allowed us to infer the contribution of cellular adaptive immunity to cross-protection, as well as the combined contributions of target cell depletion and innate immunity.

Collectively, the above findings strongly suggest that analysing real sequential infection data using mathematical models will help infer the timing and strength of host immunity, which are difficult to measure directly in laboratory experiments.
Such mathematical models will not only have the ability to explain observed experimental outcomes, but the ability to predict outcomes of new experiments, which can then be tested in the laboratory.
These findings are particularly important as sequential infection experiments are increasingly being used to study the role of the immune response during infection with influenza and other respiratory pathogens~\citep{Laurie2018,Chan2018}.

Repeating the entire study with different noisy datasets (with the same `true' parameters) did not change these findings (data and analysis not shown).

\subsection*{Limitations of sequential infection experiments}\label{sec:simulation_estimation_limitations}

This study has highlighted some limitations of quantifying the immune response using virological data from sequential infection experiments alone.

Firstly, using the synthetic sequential infection data, we could not discriminate between the effects of cellular and humoral adaptive immunity in controlling a primary infection.
If the effects of cellular and humoral adaptive immunity need to be distinguished, such as to predict the effects of vaccines that boost these components separately, quantities other than the viral load may need to be measured.

Secondly, we could not definitively rule out the possibility that target cell depletion contributed significantly to cross-protection.
We were also unable to distinguish the roles of different innate immune mechanisms in cross-protection.
Some modelling applications may require the strengths of different innate immune mechanisms to be known separately.
An example of such an application is modelling the effect of treatments that modulate the innate immune response, such as the toll-like receptor-2 agonist Pam2Cys which has been shown to stimulate innate immune signals and reduce influenza-associated mortality and morbidity in animal studies~\citep{Tan2012}.

In addition to total viral load data, the study by \Laurie{} also included infectious viral load measurements for single infection ferrets, and serological responses and cytokine levels at limited time points.
Inclusion of this data could help to address the above limitations; the utility of this additional data can be assessed by further simulation-based studies.

New experiments could also be conducted to improve parameter estimates, leading to more accurate inference of the timing and strength of immune components.
Previous studies have measured viral decay rates \textit{in vitro} and incorporated these estimates into model fitting~\citep{Pinilla2012,Paradis2015a}.
\textit{In vitro} studies can also directly measure the time course of those immune mechanisms which are active \textit{in vitro}~\citep{Mitchell2011}.

\section*{Future work and concluding summary}

Now that we have shown how mathematical models can increase the utility of sequential infection experiments, fitting the model to the experimental data by \Laurie{} is our priority.
A simulation-estimation study alone cannot validate the mathematical model used, or infer the effects of host immunity against the pathogens in the experiments. 
However, this study ensures that results will be interpreted appropriately when the models are fitted to data.

We have demonstrated that relative to single infection experiments, the sequential infection study design helps us to better understand cross-protection on short timescales.
Further, data from sequential infection experiments helps to discriminate between existing models for a primary infection, leading to an improved understanding of the control and resolution of an infection.

\section*{Materials and methods}

\subsection*{The model}\label{sec:model}

\subsubsection*{Viral dynamics}

The viral dynamics model is based on a model we previously published~\citep{Yan2017}.
It incorporates three major components of the immune response --- innate, humoral adaptive and cellular adaptive.

\begin{figure}[h!]
\begin{adjustwidth}{-2.25in}{0in}
\centering
\includegraphics[scale = .5]{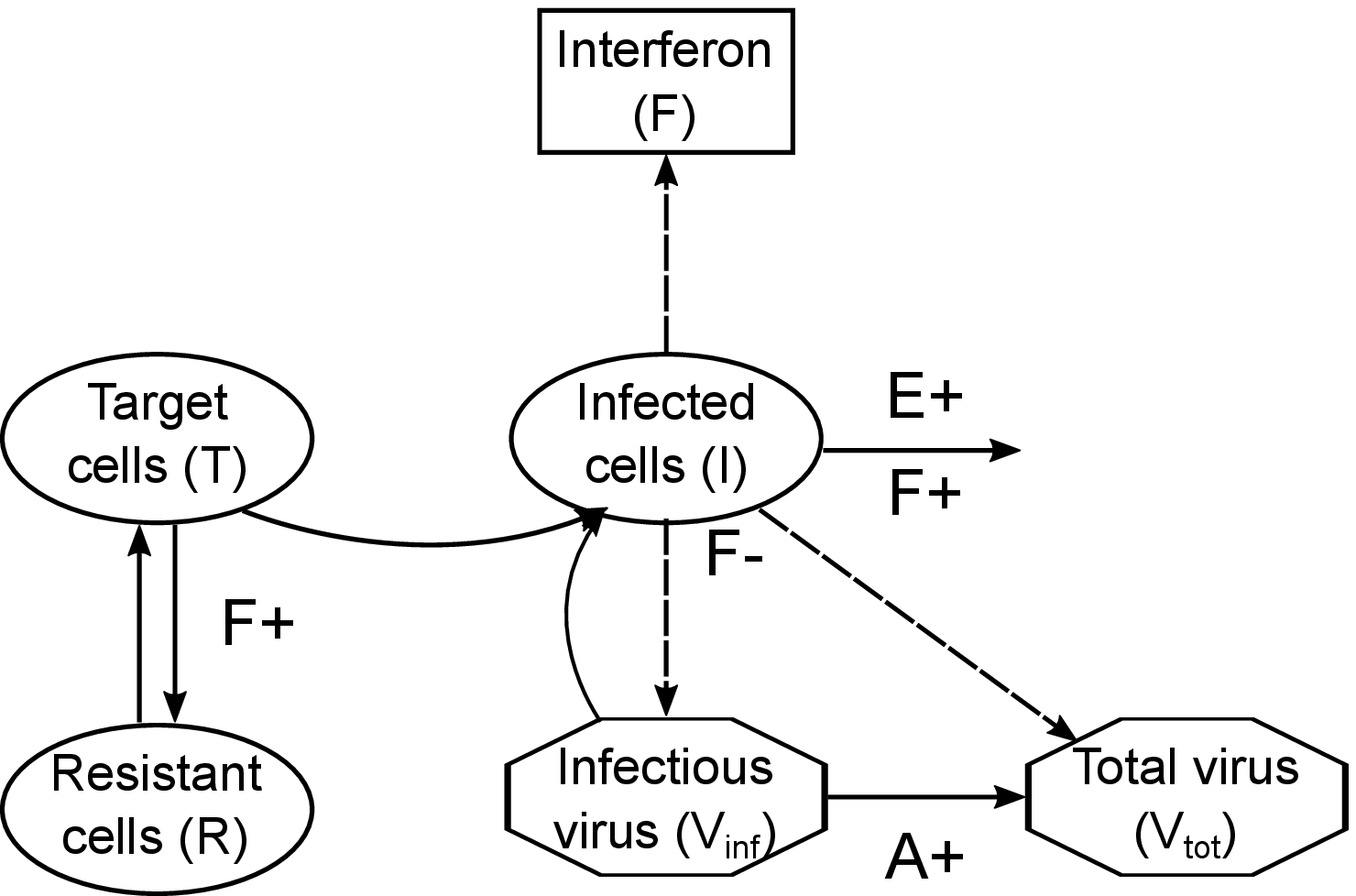}

\includegraphics[scale = .5]{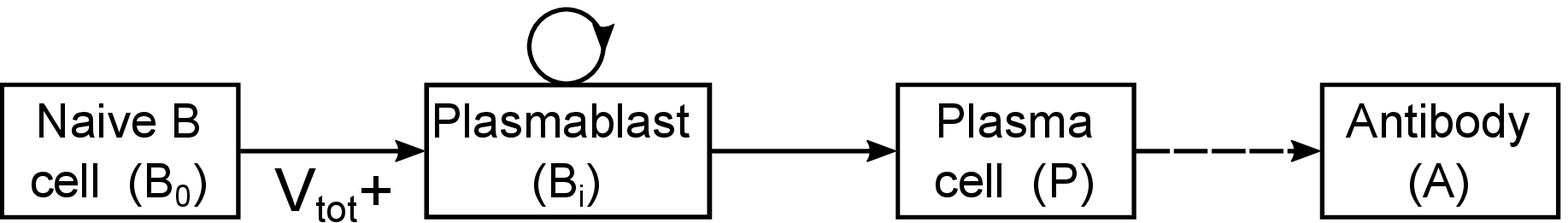}

\includegraphics[scale = .5]{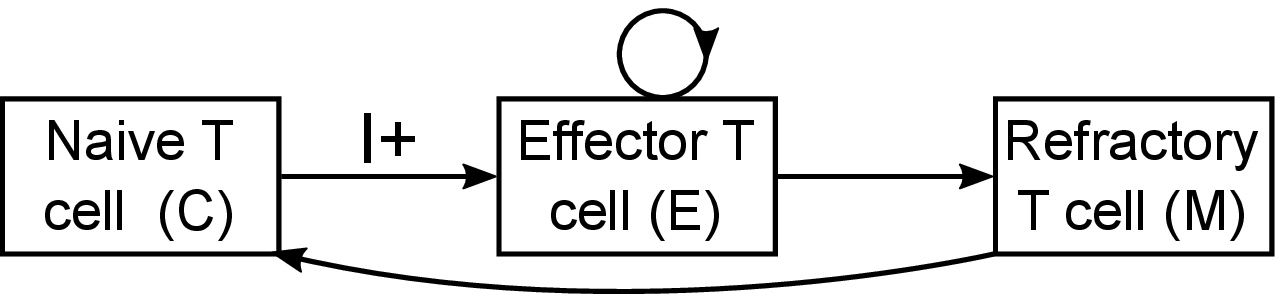}
\caption{{\bf The within-host influenza model for a single strain.}  (Top) Viral dynamics and innate immune response; (middle) humoral adaptive immune response; (bottom) cellular adaptive immune response.  Solid arrows indicate transitions between compartments or death (shown only for immune-enhanced death processes); dashed arrows indicate production; plus signs indicate an increased transition rate due to the indicated compartment.}
\label{fig:model}
\end{adjustwidth}
\end{figure}

Fig~\ref{fig:model} shows a compartmental diagram of the model for a single strain.
The system is described by a coupled set of ordinary differential equations (Eqs~\ref{eq:model}--\ref{eq:model_cellular}).

\begin{subequations}
\begin{align}
\frac{dT}{dt} &= g(T+R) \left(1-\frac{T+R+I}{T_0}\right) - \beta V_{inf} T + \rho R - \phi FT,   \label{eq:dTdt}\\
\frac{dI}{dt} &= \beta V_{inf}T - \left(\delta_{I} + \kappa_{F}F +  \kappa_{E}E\right) I, \label{eq:dIdt}\\
\frac{dV_{inf}}{dt} &= \frac{p_{Vinf}}{1+sF} I - (\delta_{Vinf}+ \kappa_{A}A + \beta T)V_{inf}, \label{eq:dVdt}\\
\frac{dV_{tot}}{dt} &= \frac{p_{Vinf}p_{Vratio}\alpha}{1+sF} I - \delta_{Vtot}V_{tot} - \alpha\beta TV_{inf}. \label{eq:dVRdt}
\end{align}
\label{eq:model}
\end{subequations}

Eq~\ref{eq:model} describes the dynamics of target cells ($T$), infected cells ($I$) and virions ($V_{inf}$ and $V_{tot}$ for infectious and total virions respectively).
Virions ($V_{inf}$) bind to target cells ($T$) to infect them; infected cells ($I$) produce virions; and infected cells and virions both decay at a constant rate.
Target cells also regrow, with an imposed carrying capacity.
Immunity is mediated by the compartments $R$ (resistant cells), $F$ (type I interferon), $A$ (antibodies) and $E$ (effector CD8$^+$ T cells), the dynamics of which will be described shortly.
Descriptions of model parameters are given in \nameref{table:parameters_replication}--\nameref{table:parameters_adaptive}, and in our previous publication~\citep{Yan2017}.

The compartment $V_{inf}$ refers to the number of infectious virions in the host; however, an infected cell produces both infectious and non-infectious virions, the latter of which arise due to defects introduced during the viral replication process~\citep{Nayak1985,Marriott2010}.  
Moreover, in the experiments conducted by \Laurie{}, the total viral load, rather than the number of infectious virions, was measured.
An additional complication is that the concentration of total nasal wash, rather than the absolute number of virions, was measured.
Hence, we include Eq~\ref{eq:dVRdt} for the total virion concentration $V_{tot}$.

Innate immunity is mediated through type I interferon ($F$), the production of which is stimulated by infected cells.
Three effects of type I interferon are modelled through Eqs~\ref{eq:model}--\ref{eq:model_innate}:

\begin{itemize}
	\item rendering target cells temporarily resistant to infection ($T \to R$);
	\item decreasing the production rate of virions from infected cells; and
	\item increasing the decay rate of infected cells.
\end{itemize}

\begin{subequations}
\begin{align}
\frac{dR}{dt} &= \phi FT - \rho R,\label{eq:dRdt}\\
\frac{dF}{dt} &= I - \delta_{F} F. \label{eq:dFdt}
\end{align}
\label{eq:model_innate}
\end{subequations}

The humoral adaptive immune response is mediated by antibodies ($A$), which bind to virions and neutralise them, rendering them non-infectious.
Naive B cells ($B_{0}$) are stimulated by virus to proliferate and differentiate into plasma cells ($P$), which produce antibodies.
Eq~\ref{eq:model_humoral} describes these processes.

\begin{subequations}
\begin{align}
\frac{dB_{0}}{dt} &= -\frac{V_{tot}}{k_{B}+V_{tot}} \beta_{B}B_{0},\label{eq:dB0dt}\\
\frac{dB_{1}}{dt} &= \frac{V_{tot}}{k_{B}+V_{tot}} \beta_{B}B_{0} - \left(\frac{n_{B}}{\tau_{B}}+ \delta_{B}\right) B_{1},\label{eq:dB1dt}\\
\frac{dB_{i}}{dt} &= \frac{2n_{B}B_{i-1}}{\tau_{B}} - \left(\frac{n_{B}}{\tau_{B}}+ \delta_{B}\right) B_{i}, \qquad i = 2, ..., n_B,\label{eq:dBidt}\\
\frac{dP}{dt} &= \frac{2n_{B}B_{n_B}}{\tau_{B}} - \delta_{B} P, \label{eq:dPdt}\\
\frac{dA}{dt} &= P - \delta_{A} A.
\end{align}
\label{eq:model_humoral}
\end{subequations}

The cellular adaptive immune response is mediated by effector CD8$^+$ T cells ($E$).
Infected cells stimulate the differentiation of effector CD8$^+$ T cells from their naive counterparts ($C$); effector CD8$^+$ T cells then increase the death rate of infected cells.
Some effector CD8$^+$ T cells remain after a primary infection as memory CD8$^+$ T cells.
After a refractory period (represented by the $M$ stage), they are modelled as having the same dynamics as naive cells, and can be re-stimulated to become effector CD8$^+$ T cells upon challenge.
Eq~\ref{eq:model_cellular} describes these processes.

\begin{subequations}
\begin{align}
\frac{dC}{dt} &= \frac{M}{\tau_{M}}-\frac{I/k_{C}}{1+I/k_{C}} \beta_{C}C,\label{eq:dCdt}\\
\frac{dE_{1}}{dt} &= \frac{I/k_{C}}{1+I/k_{C}} \beta_{C}C - \left(\frac{n_{E}}{\tau_{E}}+ \delta_{E}\right)E_{1},\label{eq:dE1dt}\\
\frac{dE_{i}}{dt} &= \frac{2n_{E}E_{i-1}}{\tau_{E}} - \left(\frac{n_{E}}{\tau_{E}}+ \delta_{E}\right) E_{i}, \qquad i = 2,...,n_E-1,\label{eq:dEidt}\\
\frac{dE_{n_{E}}}{dt} &= \frac{2n_{E}E_{n_E-1}}{\tau_{E}} - \delta_{E} E_{n_{E}}, \label{eq:dEndt}\\
E &= \sum_{i = 1}^{n_{E}}E_{i},\\
\frac{dM}{dt} &= \epsilon \delta_{E} E_{n_{E}} - \delta_{E}M - \frac{M}{\tau_{M}}\label{eq:dMdt}.
\end{align}
\label{eq:model_cellular}
\end{subequations}

When more than one strain co-infects the host, the strains interact in three ways:

\begin{itemize}
	\item competition for target cells, which become depleted due to the infection and subsequent death of cells;
	\item innate immunity, which acts across all strains; and 
	\item cellular adaptive immunity, which can be partly cross-reactive.
\end{itemize}

Activation of each of these mechanisms by the primary virus lowers the effective reproduction number of the challenge strain, but to different extents depending on parameter values.
Note that because the model includes target cell regrowth, infection with the challenge virus can become established despite target cell depletion due to the death of infected cells.
Each naive CD8$^+$ T cell pool can be stimulated by one or more virus strains, depending on model parameters; cross-reactivity arises when a T cell pool could be stimulated by more than one virus strain.
The clearance of infected cells by effector CD8$^+$ T cells is similarly strain-specific.
The antibody response is modelled as completely strain-specific, with no cross-reactivity between strains.
It is thus unnecessary to include long-term humoral adaptive immunity.
Extension of the model to include the potential effects of antibody-mediated cross-protection (as reviewed by \citet{Ekiert2012}) is the subject of future work.

\begin{figure}[!h]
\centering
\includegraphics[scale = .4]{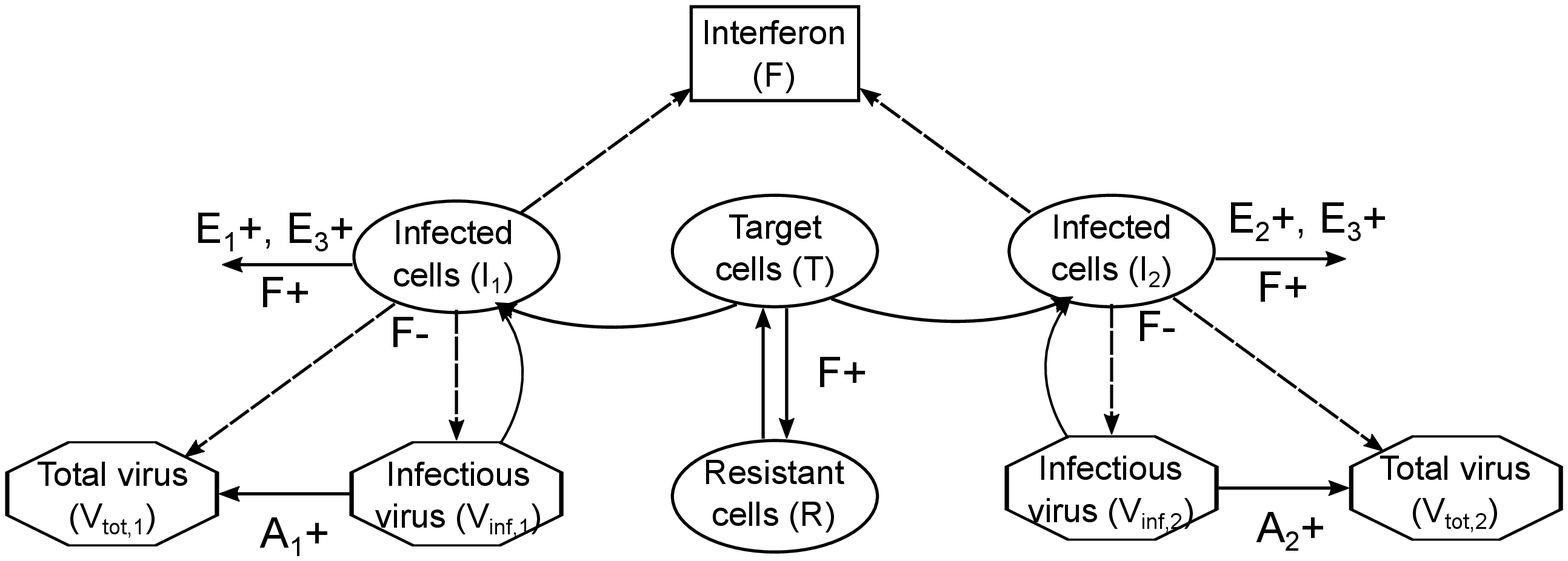}

\vspace{1cm}
\includegraphics[scale = .4]{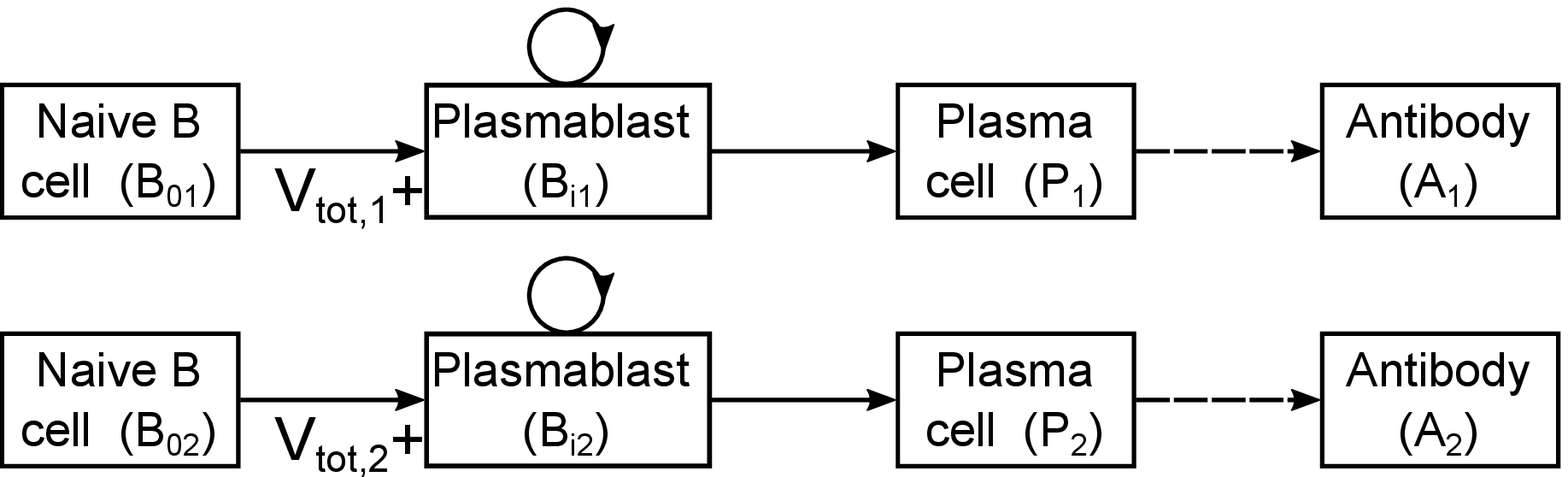}

\vspace{1cm}
\includegraphics[scale = .4]{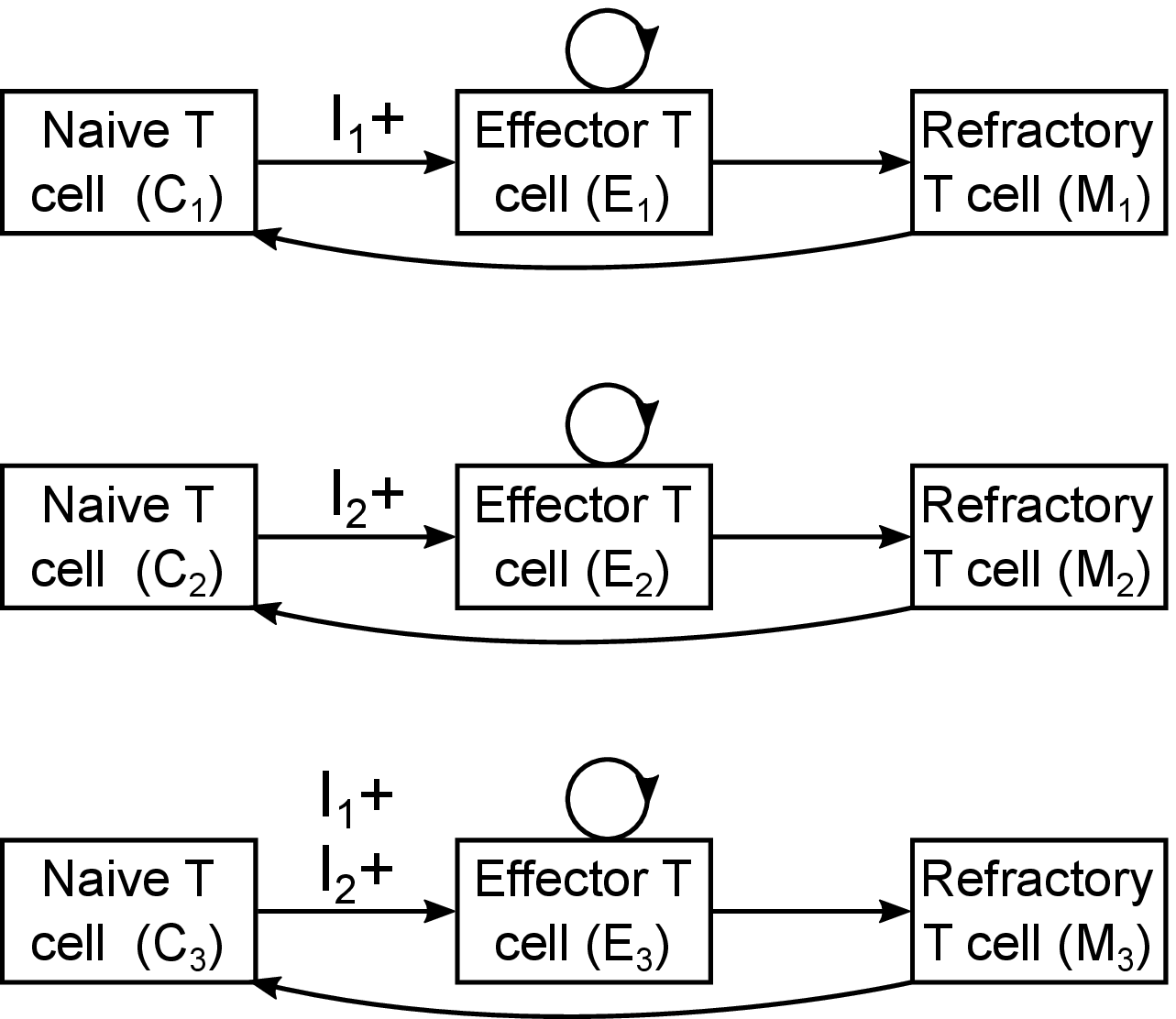}
\paragraph*{S4 Fig}
\label{fig:model_two_strain}
{\bf Compartmental diagram for two strains and three T cell pools.}
Cells infected with influenza strain 1 stimulate naive CD8$^+$ T cells in pools 1 and 3, and are cleared by effector CD8$^+$ T cells in these pools.  Cells infected with influenza strain 2 stimulate naive CD8$^+$ T cells in pools 2 and 3, and are cleared by effector CD8$^+$ T cells in these pools.
\end{figure}

\nameref{fig:model_two_strain} illustrates the model for two strains and three T cell pools; the equations are given in \nameref{file:eqs}.
Three T cell pools is a parsimonious choice, to allow for one pool to be cross-reactive between strains and two pools to be strain-specific, one for each strain.

\subsubsection*{Observation model}

Observations were simulated from the `true' viral load by adding lognormal noise and imposing a detection threshold.
Mathematically, the measured viral load $y_{qfk}$ for each virus $q = 1,2,\hdots,Q$, ferret $f = 1,2,\hdots,F$ and measuring time point $t_{qfk}$ where $k = 1,2,\hdots,K_{qf}$  is given by

\begin{subequations}
\begin{align}
y_{qfk} &= 
\begin{cases}
V_{totq}(t_{qfk},u_{f},\boldsymbol{\beta})10^{e_{qfk}} \quad\textrm{when} \quad V_{totq}(t_{qfk},u_{f},\boldsymbol{\beta})10^{e_{qfk}} \geq \Theta \\
0 \qquad \text{otherwise}
\end{cases} \\
\text{where} \qquad e_{qfk} &\overset{i.i.d.}{\sim} \mathcal{N}(0,\sigma).
\end{align}
\end{subequations}
$\boldsymbol{\beta}$ is a vector of parameter values, $u_f$ is the inter-exposure interval for ferret $f$, $e_{qfk}$ is the measurement error, and $\Theta$ is the detection threshold. 
$V_{totq}(t_{qfk},u_{f},\boldsymbol{\beta})$ is the solution to the two-strain version of Eqs~\ref{eq:model}--\ref{eq:model_cellular} for the $V_{totq}$ compartment at time $t_{qfk}$ for the given parameter values and inter-exposure intervals.
$\Theta$ takes the value 10 RNA copies/100$\mu$L in the experiments by \Laurie{}.
A measured viral load of 0 denotes that the viral load is below the detection threshold.

Therefore the likelihood of the model given the data is

\begin{subequations}
\begin{align}
P(\mathbf{y}|\boldsymbol{\theta}) &= \prod_{q=1}^Q\prod_{f=1}^F\prod_{k=1}^{K_{qf}}P(y_{qfk}|\boldsymbol{\theta}) \qquad \textrm{where}\\
P(y_{qfk}|\boldsymbol{\theta}) &= 
\begin{cases} 
\frac{1}{\sqrt{2\sigma^2\pi} } \exp \left\{-\frac{\left[\log_{10}y_{qfk}-\log_{10}V_{totq}(t_{qfk},u_{f},\boldsymbol{\beta})\right]^2}{2\sigma^2} \right\} & \text{if } y_{qfk} \geq \Theta, \\
\bigints_0^\Theta\frac{1}{\sqrt{2\sigma^2\pi} } \exp \left\{-\frac{\left[\log_{10}x-\log_{10}V_{totq}(t_{qfk},u_{f},\boldsymbol{\beta})\right]^2}{2\sigma^2} \right\} dx & \text{if } y_{qfk} = 0,\\
0 & \text{otherwise}.
\end{cases}\label{eq:likelihood}
\end{align}
\end{subequations}
In the second line of Eq~\ref{eq:likelihood}, the likelihood when the data is below the detection threshold is obtained by integrating the probability density function from 0 to the detection threshold, i.e. treating the data below the threshold as censored~\citep{Ahn2008}.
The vector $\boldsymbol{\theta}$ contains the parameters $\boldsymbol{\beta}$, the inter-exposure intervals $u_f$, the time points $t_{qfk}$, and the standard deviation $\sigma$ of the measurement error.

\subsection*{Simulated experiments}

The model and the chosen `true' parameters were used to generate synthetic data akin to that in \Laurie{}.
For six ferrets, intervals of 1, 3, 5, 7, 10 and 14 days separated exposures to two influenza strains.  
In addition, thirteen ferrets were exposed to a single influenza strain only.
The sequential infection dataset consists of the viral load for the six sequential infection ferrets and one single infection ferret; the single infection dataset consists of the viral load for the thirteen single infection ferrets.
The number of single infection ferrets was chosen such that the number of exposures to influenza virus is the same in each dataset, and so the number of data points is roughly the same.

\subsection*{Selection of model parameters to generate synthetic data}

The `true' parameter values chosen to generate the synthetic data are given in \nameref{table:parameters_replication}--\nameref{table:parameters_adaptive}.
The parameters were assumed to be identical between the two strains, except for the parameters governing cross-reactivity in the cellular adaptive immune response.
In addition to the criteria discussed in the Results section, the parameters were chosen to reproduce qualitative behaviour for a single infection when immune components are suppressed:

\begin{itemize}
\item when the innate adaptive immune response is absent ($F \to 0$), the peak viral load increases~\citep{Seo2002};
\item when the humoral adaptive immune response is absent ($A \to 0$), the viral load rebounds~\citep{Iwasaki1977};
\item when the cellular adaptive immune response is absent ($E \to 0$), resolution of the infection is delayed~\citep{Yap1978}; and
\item when both arms of the adaptive immune response are absent ($A, E \to 0$), chronic infection ensues~\citep{Kris1988}.
\end{itemize}
For an extensive evaluation of a very similar model's behaviour under these types of conditions (for single infection events), see \citet{Cao2016}.

In addition to parameter values, initial values were required when simulating infections.
For a single infection, the initial values for all compartments in Eqs~\ref{eq:model}--\ref{eq:model_cellular} except $T$, $V_{inf}$, $V_{tot}$, $C$ and $B_0$ were zero.
The initial values of $C$ and $B_0$ (naive T and B cells respectively) were normalised to 1.
The initial values of $T$ and $V_{inf}$ (the number of target cells and the concentration of infectious virions respectively) were estimated parameters.
The initial concentration of total virus was then $V_{tot}(0) = \gamma \alpha V_{inf}(0)$, where $\gamma$ and $\alpha$ were conversion parameters described in \nameref{table:parameters_replication}.
For sequential infections, the conditions at the time of the primary exposure were as above; the system was integrated until the time of the challenge exposure, at which $V_{inf,2}(0)$ infectious virions for the challenge strain was added to the system, and the total concentration of the challenge strain was set to $V_{tot,2}(0)$.

\subsection*{Model fitting}

\subsubsection*{Parameters to be estimated}

All model parameters were estimated, except for the following parameters which were either fixed or a function of other estimated parameters.
We fixed two parameters --- the number of plasmablast division cycles ($n_B$) and the number of effector CD8$^+$ T cell division cycles ($n_E$) --- to be 5~\citep{Marchuk1991,Sze2000} and 20~\citep{VanStipdonk2001} respectively.
In addition, when fitting the model to single infection data, we considered the optimistic scenario where we had independent, perfect information about the proportion of cellular adaptive immunity during a primary infection that was cross-reactive with the challenge strain.
As one T cell pool was cross-reactive between strains and two pools were strain-specific, this amounted to fixing the proportion of cellular adaptive immunity attributed to each T cell pool.
We did so by fixing the numbers of infected cells for half-maximal stimulation of naive CD8$^+$ T cells $k_{Cj1}$ to their `true' values for each T cell pool $j$.
Then when we extended the model to two strains, we set $k_{Cjq}$ to these same `true' values.
We then calculated the scaled clearance rates of infected cells by effector CD8$^+$ T cells $\kappa_{Ejq}$ by taking the fitted value of $\kappa_{E11}$, and applying the formula $\kappa_{Ejq} = \kappa_{E11}k_{C11}/{k_{Cjq}}$.

Instead of fitting the infectivity ($\beta$) and the production rate of infectious virions from an infected cell ($p_{Vinf}$), we fitted the basic reproduction number $R_0$ (Eq~\ref{eq:R_0}) and the initial viral load growth rate $r$ (Eq~\ref{eq:r}), as we hypothesised that these were more intimately linked to features of the viral load curve.
Practically speaking, we proposed a new value for $R_0$ (or $r$), calculated the corresponding values of $\beta$ and $p_{Vinf}$, solved the model equations, calculated the likelihood of the data given the parameters, and accepted or rejected the new value for $R_0$ (or $r$).

The basic reproduction number $R_0$ is the mean number of secondary infected cells due to (the virions produced by) a single infected cell.
The expression for $R_0$ is

\begin{equation}
R_0 = \frac{\beta T_0 p_{Vinf}}{(\delta_{Vinf} + \beta T_0)\delta_I},
\label{eq:R_0}
\end{equation}
and is the same as that for a model without a time-dependent immune response \citep{Beauchemin2008}.

The viral load during early infection can be approximated by

\begin{equation}
V = V_0\exp(rt).
\end{equation}
Arenas~\textit{et al.}~\citet{Arenas2017} showed using a simulation-estimation study that this parameter was well estimated even when only viral load data was available.

The expression for $r$, derived by linearising Eq~\ref{eq:model} around the disease-free equilibrium~\citep{Nowak1997}, is

\begin{equation}
r = -\frac{\delta_{Vinf} + \beta T_0 + \delta_I}{2} + \frac{\sqrt{(\delta_I - \delta_{Vinf} - \beta T_0)^2 + 4 \beta T_0 p_{Vinf}}}{2}.
\label{eq:r}
\end{equation}

\subsubsection*{Prior distributions}

We began with a uniform distribution in parameter space whose bounds along each dimension are given in \nameref{table:parameters_replication}--\nameref{table:parameters_adaptive}.  
Note that parameter estimation was performed in a parameter space where all parameters except for the standard deviation of the measurement error $\sigma$ were log transformed.
Then, we excluded regions of parameter space where the parameters $\log_{10}\beta$ and $\log_{10}p_{Vinf}$, which were not directly estimated but were instead recovered from Eqs~\ref{eq:R_0} and~\ref{eq:r}, were outside the bounds given in \nameref{table:parameters_replication}.

The priors were deliberately chosen to be wide because previous parameter estimates came from a range of experimental systems, and parameters with similar physical definitions could vary in value depending on the model used.
The bounds for viral replication parameters were based on those by Petrie \textit{et al.}~\citet{Petrie2015} where the equivalent parameters exist.
Otherwise, where multiple estimates existed in the literature (as cited in the tables), the bounds were chosen to encompass all of them.  
Where we could only find a single estimate, bounds spanning at least an order of magnitude were chosen (unless the parameter is a pure rate parameter, as discussed shortly).
Where no estimate was found, we assigned very wide bounds spanning much more than one order of magnitude.
In general, the bounds for pure rate parameters (those with units day$^{-1}$ only) were chosen to be narrower as their order of magnitude was known, whereas bounds for parameters such as $R_0$ were much wider.

Furthermore, for computational efficiency, some minimal constraints on the behaviour of the viral load and timing of various immune components were incorporated into the prior distribution.  
These constraints were imposed because parameter sets that generate `unreasonable' viral load trajectories for a single infection caused large delays in numerical integration of the two-strain differential equations.  
The inclusion criteria were that for a single infection,

\begin{itemize}
\item the total viral load rises by at least one order of magnitude during infection;
\item the total viral load peaks 0--7 days post-exposure;
\item the humoral adaptive immune response is not active too early (five days post-exposure, the neutralisation rate of virus by antibodies, $\kappa_{A} A$, does not exceed $10^3$ day$^{-1}$); and
\item the cellular adaptive immune response is not active too early (five days post-exposure, the clearance rate of infected cells by effector CD8$^+$ T cells, $\sum_{j = 1}^J \kappa_{Ej1} E_j$, does not exceed $10^3$ day$^{-1}$).
\end{itemize}

If the viral load trajectory (in the absence of measurement noise) predicted by a parameter set does not fulfil all of these conditions, the value of the prior distribution is zero at that point in parameter space.  

\subsubsection*{Model fitting algorithm}

We fitted the model using the Metropolis algorithm~\citep{Metropolis1949,Metropolis1953} embedded within a Gibbs sampler structure~\citep{Geman1984}, implemented in Octave 3.8.2~\citep{Octave}.
To evaluate the likelihood, Eqs~\ref{eq:model}--\ref{eq:model_cellular} were solved using the CVODE solvers developed by \citet{Cohen1996}, implemented in MATLAB~\citep{Vanlier2012a}.
Of the available solvers, a backward differentiation formula method in variable order, variable step, fixed leading coefficient form was chosen.
Extinction was enforced by defining an infection to have resolved if both the number of infected cells and virions was below 0.1.

To assess convergence, three chains were run in parallel using different starting parameter values $\boldsymbol{\theta}_0$ drawn from  the prior distribution.
The procedure for determining the number of iterations for which to run the chains is detailed in \nameref{text:more_fitting}.
For efficient mixing, the proposal distributions were tuned such that the proportion of accepted proposals was not too low or too high, as detailed in \nameref{text:more_fitting}.
For each of the three chains, parallel tempering (as developed by \citet{Geyer1991} and reviewed by \citet{Earl2005}) was implemented to improve exploration of parameter space.
The number of iterations before testing whether to swap chains in the parallel tempering process was set to 10.
During the calibration period for the proposal distributions, the temperatures were also calibrated~\citep{Kone2005}, as detailed in \nameref{text:more_fitting}.
Once convergence was reached, the effective sample size was calculated for each chain (using the iterations that were kept following the burn-in process) using the \texttt{effectiveSize} function in the \texttt{coda}~\citep{coda} package in R~\citep{R}.
Convergence diagnostics for the chains are shown in \nameref{file:diagnostics}.

The marginal posterior distributions in this study are plotted in \nameref{file:marginals} using all samples from the chains (after burn-in), without thinning.
When using the joint posterior distribution to make predictions, we used $10^4$ parameter sets corresponding to uniformly spaced iterations in each of the chains.

Results were visualised using MATLAB R2015b~\citep{MATLAB}.

\subsection*{Model predictions}

First, to determine whether the fitted model captured the timing and strength of each immune component during a primary infection, we used parameter sets from the joint posterior distribution to simulate the viral load during a single infection, using a modified model where either

\begin{itemize}
	\item adaptive immunity is suppressed;
	\item innate and adaptive immunity is suppressed;
	\item cellular adaptive immunity is suppressed; or
	\item humoral adaptive immunity is suppressed.
\end{itemize}
95\% prediction intervals were constructed using these simulations.

Second, to determine whether the fitted model captured cross-protection between strains, we used parameter sets from the joint posterior distribution to simulate different inter-exposure intervals.

Third, to determine whether the fitted model captured the contribution of each immune component to cross-protection between strains, we used parameter sets from the joint posterior distribution to simulate the viral load during sequential infection, using a modified model where either

\begin{itemize}
	\item cross-protection is only mediated by cellular adaptive immunity, and not target cell depletion or innate immunity (model XC); 
	\item cross-protection is mediated by innate immunity, but not target cell depletion or cellular adaptive immunity (model XI); or
	\item cross-protection is mediated by target cell depletion and/or innate immunity, but not cellular adaptive immunity (model XIT).
\end{itemize}

Details of models XC, XI and XIT are provided in \nameref{file:eqs}.

Table~\ref{table:modified_models} summarises the model modifications in this section.

\begin{table}[!ht]
\begin{adjustwidth}{-2.25in}{0in} 
\centering
\caption{
{\bf Summary of model modifications for predictions.}}
\begin{tabular}{|l|l|l|l|l|}
\hline
Model & Target cells & Interferon & Antibodies & T cells \\ \thickhline
Baseline & shared & shared & separate & partly shared\\ \hline
No adaptive immunity & shared & shared & none & none \\ \hline
No immunity & shared & none & none & none\\ \hline
No cellular adaptive immunity & shared & shared & separate & none \\ \hline
No humoral adaptive immunity & shared & shared & none & partly shared\\ \hline
XC & separate & separate & separate & partly shared\\ \hline
XI & separate & shared & separate & separate\\ \hline
XIT & shared & shared & separate & separate\\ \hline
\end{tabular}
\begin{flushleft}
`Shared' denotes that the compartment in the table header interacts with all virus strains.
For example, if interferon are `shared', all virus strains induce production of the same interferon, and interferon's antiviral effects apply to all strains.
`Separate' denotes that the compartment interacts with one virus strain only.
For example, if interferon are `separate', each virus strain induces the production of a separate pool of interferon, and the antiviral effects of each pool of interferon apply only to that strain.
T cells being `partly shared' denotes that some T cell pools interact with one virus strain only, while other T cell pools are stimulated by more than one strain and clear cells infected by any of those strains.
`None' denotes that the compartment is removed from the model.
\end{flushleft}
\label{table:modified_models}
\end{adjustwidth}
\end{table}

\clearpage

\section*{Supporting information}


\clearpage
\paragraph*{S1 Text.}
\label{text:more_fitting}
{\bf More details on the model fitting procedure.}

\paragraph*{S2 Text.}
\label{text:notes_V_tot_values}
{\bf Notes on biologically plausible ranges for the parameters $p_{Vratio}$, $\alpha$ and $\gamma$, as provided in \nameref{table:parameters_replication}.}

\paragraph*{S1 File.}
\label{file:eqs}
{\bf Two-strain model equations for the baseline and modified models, and compartmental diagrams for the modified models.}
In the baseline model, naive CD8$^+$ T cell pools 1 and 3 are stimulated by cells infected with strain 1, and naive CD8$^+$ T cell pools 2 and 3 are stimulated by cells infected with strain 2.
The clearance of infected cells by effector CD8$^+$ T cells is similarly strain-specific.  

\paragraph*{S2 File.}
\label{file:high_xreact}
{\bf Results for an additional set of parameters where the degree of cross-reactivity in the cellular adaptive immune response is high.}  

\paragraph*{S3 File.}
\label{file:diagnostics}
{\bf Convergence diagnostics for the MCMC chains.}  

\paragraph*{S4 File.}
\label{file:marginals}
{\bf Marginal posterior distributions for the model parameters.}  

\paragraph*{S1 Table.}
\label{table:parameters_replication}
{\bf Viral replication parameter values and prior bounds.}
Note that the values and prior bounds are given in logarithmic space.  For example, the value of $\log_{10} R_0$ was $\log_{10}4.9$ and the prior bounds were $[0,3]$.  Hence, the value of $R_0$ was 4.9 and the prior bounds of $R_0$ were $[1, 1000]$.  
$\beta$ and $p_{Vinf}$ were not directly fitted, but their values as recovered from Eqs.~\ref{eq:R_0} and~\ref{eq:r} could not exceed the bounds given.  
Because total virions include infectious virions, the total virion decay rate should be slower than the infectious virion decay rate.  Hence, the difference between the infectious and total virion decay rates $\delta_{Vinf}-\delta_{Vtot}$, rather than the infectious virion decay rate $\delta_{Vinf}$, was fitted to ensure that the former quantity was positive.
Notes on biologically plausible ranges for the parameters $p_{tot}$, $\alpha$ and $\gamma$ are given in \nameref{text:notes_V_tot_values}.

\paragraph*{S2 Table.}
\label{table:parameters_innate}
{\bf Innate immune response parameter values and prior bounds.}

\paragraph*{S3 Table.}
\label{table:parameters_cross-react}
{\bf Values and prior bounds for the cross-reactivity parameters in the cellular adaptive immune response.}
The number of infected cells for half-maximal stimulation of naive/memory CD8$^+$ T cells $k_{Cjq}$ and the (scaled) clearance rate of infected cells by effector CD8$^+$ T cells $\kappa_{E11}$.

\paragraph*{S4 Table.}
\label{table:parameters_adaptive}
{\bf Adaptive immune response and observation model parameter values and prior bounds.}

\section*{Acknowledgments}
We thank Karen~L.~Laurie for designing and performing the experiments modelled, and providing virological insight.
We also acknowledge helpful discussions with Patricia~T.~Campbell, Pengxing~Cao, Steven~Riley and Alexander~E.~Zarebski.

\section*{Author contributions}

\begin{itemize}
\item Conceptualisation: AWCY, SGZ, JAS, JMM.
\item Formal analysis: AWCY.
\item Funding acquisition: JAS, JMM.
\item Investigation: AWCY.
\item Methodology: AWCY, SGZ, JMM.
\item Project administration: JMM.
\item Software: AWCY.
\item Supervision: SGZ, JMM.
\item Validation: AWCY.
\item Visualisation: AWCY.
\item Writing -- Original draft preparation: AWCY, JMM.
\item Writing -- review and editing: AWCY, SGZ, JAS, JMM.
\end{itemize}

\nolinenumbers

%
%
%





\bibliography{library_checked,unpublished}

\begin{thebibliography}{10}

\bibitem{Dobrovolny2013}
Dobrovolny HM, Reddy MB, Kamal MA, Rayner CR, Beauchemin CAA.
\newblock {Assessing mathematical models of influenza infections using features
  of the immune response.}
\newblock PLoS ONE. 2013;8(2):e57088.
\newblock doi:{10.1371/journal.pone.0057088}.

\bibitem{Seo2002}
Seo SH, Hoffmann E, Webster RG.
\newblock {Lethal H5N1 influenza viruses escape host anti-viral cytokine
  responses.}
\newblock Nat Med. 2002;8(9):950--4.
\newblock doi:{10.1038/nm757}.

\bibitem{Iwasaki1977}
Iwasaki T, Nozima T.
\newblock {Defense mechanisms against primary influenza virus infection in mice
  I. The roles of interferon and neutralizing antibodies and thymus dependence
  of interferon and antibody production}.
\newblock J Immunol. 1977;118(1):256--263.

\bibitem{Yap1978}
Yap KL, Ada GL.
\newblock {Cytotoxic T cells in the lungs of mice infected with an influenza A
  virus.}
\newblock Scand J Immunol. 1978;7(1):73--80.
\newblock doi:{10.1111/j.1365-3083.1978.tb00428.x}.

\bibitem{Kris1988}
Kris RM, Yetter RA, Cogliano R, Ramphal R, Small PA.
\newblock {Passive serum antibody causes temporary recovery from influenza
  virus infection of the nose, trachea and lung of nude mice.}
\newblock Immunology. 1988;63(3):349--53.

\bibitem{Baccam2006}
Baccam P, Beauchemin C, Macken CA, Hayden FG, Perelson AS.
\newblock {Kinetics of influenza A virus infection in humans}.
\newblock J Virol. 2006;80(15):7590--7599.
\newblock doi:{10.1128/JVI.01623-05}.

\bibitem{Handel2010}
Handel A, Longini~Jr IM, Antia R.
\newblock {Towards a quantitative understanding of the within-host dynamics of
  influenza A infections}.
\newblock J R Soc Interface. 2010;7(42):35--47.
\newblock doi:{10.1098/rsif.2009.0067}.

\bibitem{Pawelek2012}
Pawelek KA, Huynh GT, Quinlivan M, Cullinane A, Rong L, Perelson AS.
\newblock {Modeling within-host dynamics of influenza virus infection including
  immune responses}.
\newblock PLoS Comput Biol. 2012;8(6):e1002588.
\newblock doi:{10.1371/journal.pcbi.1002588}.

\bibitem{Saenz2010}
Saenz RA, Quinlivan M, Elton D, MacRae S, Blunden AS, Mumford JA, et~al.
\newblock Dynamics of influenza virus infection and pathology.
\newblock J Virol. 2010;84(8):3974--3983.
\newblock doi:{10.1128/JVI.02078-09}.

\bibitem{Hancioglu2007}
Hancioglu B, Swigon D, Clermont G.
\newblock {A dynamical model of human immune response to influenza A virus
  infection}.
\newblock J Theor Biol. 2007;246(1):70--86.
\newblock doi:{10.1016/j.jtbi.2006.12.015}.

\bibitem{Bocharov1994}
Bocharov GA, Romanyukha AA.
\newblock {Mathematical model of antiviral immune response III. Influenza A
  virus infection}.
\newblock J Theor Biol. 1994;167(4):323--360.

\bibitem{Miao2010}
Miao H, Hollenbaugh JA, Zand MS, Holden-Wiltse J, Mosmann TR, Perelson AS,
  et~al.
\newblock {Quantifying the early immune response and adaptive immune response
  kinetics in mice infected with influenza A virus}.
\newblock J Virol. 2010;84(13):6687--6698.
\newblock doi:{10.1128/JVI.00266-10}.

\bibitem{Lee2009a}
Lee HY, Topham DJ, Park SY, Hollenbaugh J, Treanor J, Mosmann TR, et~al.
\newblock {Simulation and prediction of the adaptive immune response to
  influenza A virus infection}.
\newblock J Virol. 2009;83(14):7151--7165.
\newblock doi:{10.1128/JVI.00098-09}.

\bibitem{Li2014}
Li Y, Handel A.
\newblock {Modeling inoculum dose dependent patterns of acute virus
  infections.}
\newblock J Theor Biol. 2014;347:63--73.
\newblock doi:{10.1016/j.jtbi.2014.01.008}.

\bibitem{Ahmed2017}
{Ahmed} H, {Moore} J, {Manicassamy} B, {Garcia-Sastre} A, {Handel} A, {Antia}
  R.
\newblock {Mathematical analysis of a mouse experiment suggests little role for
  resource depletion in controlling influenza infection within host}.
\newblock ArXiv e-prints. 2017;.

\bibitem{Laurie2015}
Laurie KL, Guarnaccia TA, Carolan LA, Yan AWC, Aban M, Petrie S, et~al.
\newblock {Interval between infections and viral hierarchy are determinants of
  viral interference following influenza virus infection in a ferret model}.
\newblock J Infect Dis. 2015;212(11):1701--1710.
\newblock doi:{10.1093/infdis/jiv260}.

\bibitem{Laurie2018}
Laurie KL, Horman W, Carolan LA, Chan KF, Layton D, Bean A, et~al.
\newblock Evidence for viral interference and cross-reactive protective
  immunity between influenza B virus lineages.
\newblock J Infect Dis. 2018;217(4):548--559.
\newblock doi:{10.1093/infdis/jix509}.

\bibitem{Carolan2016}
Carolan LA, Rockman S, Borg K, Guarnaccia T, Reading P, Mosse J, et~al.
\newblock Characterization of the localized immune response in the respiratory
  tract of ferrets following infection with influenza A and B viruses.
\newblock J Virol. 2016;90(6):2838--2848.
\newblock doi:{10.1128/JVI.02797-15}.

\bibitem{Cao2015}
Cao P, Yan AWC, Heffernan JM, Petrie S, Moss RG, Carolan LA, et~al.
\newblock {Innate immunity and the inter-exposure interval determine the
  dynamics of secondary influenza virus infection and explain observed viral
  hierarchies}.
\newblock PLoS Comput Biol. 2015;11(8):e1004334.
\newblock doi:{10.1371/journal.pcbi.1004334}.

\bibitem{Yan2016}
Yan AWC, Cao P, McCaw JM.
\newblock {On the extinction probability in models of within-host infection:
  the role of latency and immunity}.
\newblock J Math Biol. 2016;73(4):787--813.
\newblock doi:{10.1007/s00285-015-0961-5}.

\bibitem{Cao2016}
Cao P, Wang Z, Yan AWC, McVernon J, Xu J, Heffernan JM, et~al.
\newblock {On the role of CD8+ T cells in determining recovery time from
  influenza virus infection}.
\newblock Front Immunol. 2016;7:611.
\newblock doi:{10.3389/fimmu.2016.00611}.

\bibitem{Yan2017}
Yan AWC, Cao P, Heffernan JM, McVernon J, Quinn KM, {La Gruta} NL, et~al.
\newblock {Modelling cross-reactivity and memory in the cellular adaptive
  immune response to influenza infection in the host}.
\newblock J Theor Biol. 2017;413:34--49.
\newblock doi:{10.1016/j.jtbi.2016.11.008}.

\bibitem{Chan2018}
Chan KF, Carolan LA, Korenkov D, Druce J, McCaw J, Reading PC, et~al.
\newblock Investigating Viral Interference Between Influenza A Virus and Human
  Respiratory Syncytial Virus in a Ferret Model of Infection.
\newblock The Journal of Infectious Diseases. 2018; p. jiy184.
\newblock doi:{10.1093/infdis/jiy184}.

\bibitem{Tan2012}
Tan ACL, Mifsud EJ, Zeng W, Edenborough K, McVernon J, Brown LE, et~al.
\newblock {Intranasal administration of the TLR2 agonist Pam2Cys provides rapid
  protection against influenza in mice}.
\newblock Molecular Pharmaceutics. 2012;9(9):2710--2718.
\newblock doi:{10.1021/mp300257x}.

\bibitem{Pinilla2012}
Pinilla LT, Holder BP, Abed Y, Boivin G, Beauchemin CAA.
\newblock {The H275Y neuraminidase mutation of the pandemic A/H1N1 influenza
  virus lengthens the eclipse phase and reduces viral output of infected cells,
  potentially compromising fitness in ferrets}.
\newblock J Virol. 2012;86(19):10651--10660.
\newblock doi:{10.1128/JVI.07244-11}.

\bibitem{Paradis2015a}
Paradis EG, Pinilla LT, Holder BP, Abed Y, Boivin G, Beauchemin CAA.
\newblock {Impact of the H275Y and I223V mutations in the neuraminidase of the
  2009 pandemic influenza virus in vitro and evaluating experimental
  reproducibility}.
\newblock PLoS ONE. 2015;10(5):e0126115.
\newblock doi:{10.1371/journal.pone.0126115}.

\bibitem{Mitchell2011}
Mitchell H, Levin D, Forrest S, Beauchemin CAA, Tipper J, Knight J, et~al.
\newblock {Higher level of replication efficiency of 2009 (H1N1) pandemic
  influenza virus than those of seasonal and avian strains: kinetics from
  epithelial cell culture and computational modeling.}
\newblock J Virol. 2011;85(2):1125--35.
\newblock doi:{10.1128/JVI.01722-10}.

\bibitem{Nayak1985}
Nayak DP, Chambers TM, Akkina RK.
\newblock In: Cooper M, Eisen H, Goebel W, Hofschneider PH, Koprowski H,
  Melchers F, et~al., editors. Defective-interfering (DI) RNAs of influenza
  viruses: origin, atructure, expression, and interference. Berlin, Heidelberg:
  Springer Berlin Heidelberg; 1985. p. 103--151.

\bibitem{Marriott2010}
Marriott AC, Dimmock NJ.
\newblock {Defective interfering viruses and their potential as antiviral
  agents}.
\newblock Rev Med Virol. 2010;20(1):51--62.
\newblock doi:{10.1002/rmv.641}.

\bibitem{Ekiert2012}
Ekiert DC, Wilson IA.
\newblock {Broadly neutralizing antibodies against influenza virus and
  prospects for universal therapies}.
\newblock Current Opinion in Virology. 2012;2(2):134--141.
\newblock doi:{10.1016/j.coviro.2012.02.005}.

\bibitem{Ahn2008}
Ahn JE, Karlsson MO, Dunne A, Ludden TM.
\newblock {Likelihood based approaches to handling data below the
  quantification limit using NONMEM VI}.
\newblock J Pharmacokinet Pharmacodyn. 2008;35(4):401--421.
\newblock doi:{10.1007/s10928-008-9094-4}.

\bibitem{Marchuk1991}
Marchuk GI, Petrov RV, Romanyukha AA, Bocharov GA.
\newblock {Mathematical model of antiviral immune response. I. Data analysis,
  generalized picture construction and parameters evaluation for hepatitis B.}
\newblock J Theor Biol. 1991;151(1):1--40.
\newblock doi:{10.1016/S0022-5193(05)80142-0}.

\bibitem{Sze2000}
Sze DMY, Toellner KM, {Garc{\'{i}}a de Vinuesa} C, Taylor DR, MacLennan ICM.
\newblock {Intrinsic constraint on plasmablast growth and extrinsic limits of
  plasma cell survival.}
\newblock J Exp Med. 2000;192(6):813--821.
\newblock doi:{10.1084/jem.192.6.813}.

\bibitem{VanStipdonk2001}
van Stipdonk MJB, Lemmens EE, Schoenberger SP.
\newblock {Na{\"{i}}ve CTLs require a single brief period of antigenic
  stimulation for clonal expansion and differentiation.}
\newblock Nat Immunol. 2001;2(5):423--429.
\newblock doi:{10.1038/87730}.

\bibitem{Beauchemin2008}
Beauchemin CAA, McSharry JJ, Drusano GL, Nguyen JT, Went GT, Ribeiro RM, et~al.
\newblock {Modeling amantadine treatment of influenza A virus in vitro}.
\newblock J Theor Biol. 2008;254(2):439--451.
\newblock doi:{10.1016/j.jtbi.2008.05.031}.

\bibitem{Arenas2017}
Arenas AR, Thackar NB, Haskell EC.
\newblock {The logistic growth model as an approximating model for viral load
  measurements of influenza A virus}.
\newblock Math Comput Simul. 2017;133:206--222.
\newblock doi:{10.1016/j.matcom.2016.10.002}.

\bibitem{Nowak1997}
Nowak MA, Lloyd AL, Vasquez GM, Wiltrout TA, Wahl LM, Bischofberger N, et~al.
\newblock {Viral dynamics of primary viremia and antiretroviral therapy in
  simian immunodeficiency virus infection.}
\newblock J Virol. 1997;71(10):7518--25.

\bibitem{Petrie2015}
Petrie SM, Butler J, Barr IG, McVernon J, Hurt AC, McCaw JM.
\newblock {Quantifying relative within-host replication fitness in influenza
  virus competition experiments}.
\newblock J Theor Biol. 2015;382:259--271.
\newblock doi:{10.1016/j.jtbi.2015.07.003}.

\bibitem{Metropolis1949}
Metropolis N, Ulam S.
\newblock {The Monte Carlo method}.
\newblock JASA. 1949;44(247):335--341.
\newblock doi:{10.1080/01621459.1949.10483310}.

\bibitem{Metropolis1953}
Metropolis N, Rosenbluth AW, Rosenbluth MN, Teller AH, Teller E.
\newblock {Equation of state calculations by fast computing machines}.
\newblock J Chem Phys. 1953;21(6):1087--1092.
\newblock doi:{10.1063/1.1699114}.

\bibitem{Geman1984}
Geman S, Geman D.
\newblock {Stochastic relaxation, Gibbs distributions, and the Bayesian
  restoration of images}.
\newblock IEEE Trans Pattern Anal Mach Intell. 1984;PAMI-6(6):721--741.
\newblock doi:{10.1109/TPAMI.1984.4767596}.

\bibitem{Octave}
John W~Eaton SH David~Bateman, Wehbring R.
\newblock {GNU Octave} version 3.8.1 manual: a high-level interactive language
  for numerical computations.
\newblock CreateSpace Independent Publishing Platform; 2014.
\newblock Available from:
  \url{http://www.gnu.org/software/octave/doc/interpreter}.

\bibitem{Cohen1996}
Cohen SD, Hindmarsh AC, Dubois PF.
\newblock {CVODE, a stiff/nonstiff ODE solver in C}.
\newblock Computers in Physics. 1996;10:138--148.
\newblock doi:{10.1063/1.4822377}.

\bibitem{Vanlier2012a}
Vanlier J, Tiemann CA, Hilbers PAJ, van Riel NAW.
\newblock {A Bayesian approach to targeted experiment design}.
\newblock Bioinformatics. 2012;28(8):1136--1142.
\newblock doi:{10.1093/bioinformatics/bts092}.

\bibitem{Geyer1991}
Geyer CJ.
\newblock {Markov chain Monte Carlo maximum likelihood}.
\newblock In: Computing Science and Statistics: Proceedings of the 23rd
  Symposium on the Interface. Interface Foundation of North America; 1991. p.
  156.

\bibitem{Earl2005}
Earl DJ, Deem MW.
\newblock {Parallel tempering: theory, applications, and new perspectives}.
\newblock Phys Chem Chem Phys. 2005;7:3910--3916.
\newblock doi:{10.1039/B509983H}.

\bibitem{Kone2005}
Kone A, Kofke DA.
\newblock {Selection of temperature intervals for parallel-tempering
  simulations}.
\newblock J Chem Phys. 2005;122(20):206101.
\newblock doi:{10.1063/1.1917749}.

\bibitem{coda}
Plummer M, Best N, Cowles K, Vines K.
\newblock {CODA: convergence diagnosis and output analysis for MCMC}.
\newblock R News. 2006;6(1):7--11.

\bibitem{R}
{R Core Team}. R: A Language and Environment for Statistical Computing; 2016.
\newblock Available from: \url{https://www.R-project.org/}.

\bibitem{MATLAB}
MATLAB.
\newblock R2015b.
\newblock Natick, Massachusetts: The MathWorks Inc.; 2015.

\end{thebibliography}

\end{document}